\documentclass[aps,prb,twocolumn,superscriptaddress]{revtex4-1}
\usepackage[breaklinks=true,colorlinks,citecolor=blue,linkcolor=blue,urlcolor=blue]{hyperref}
\usepackage{amsmath}
\usepackage{amsfonts}
\usepackage{amssymb}
\usepackage{graphicx}
\begin{document}

\title{Aspects of symmetry and topology in the charge density wave phase of 1\textit{T}-TiSe$_2$}
\author{Shin-Ming Huang}
\email{shinming@mail.nsysu.edu.tw}
\affiliation{Department of Physics, National Sun Yat-Sen University, Kaohsiung 80424, Taiwan}

\author{Su-Yang~Xu}
\affiliation{Department of Physics, Massachusetts Institute of Technology, Cambridge, Massachusetts 02139, USA}

\author{Bahadur Singh}
\affiliation{Department of Condensed Matter Physics and Materials Science, Tata Institute of Fundamental Research, Colaba, Mumbai 400005, India}

\author{Ming-Chien Hsu}
\affiliation{Department of Physics, National Sun Yat-Sen University, Kaohsiung 80424, Taiwan}

\author{Chuang-Han Hsu}
\affiliation {Centre for Advanced 2D Materials and Graphene Research Centre, National University of Singapore, 6 Science Drive 2, Singapore 117546}
\affiliation {Department of Physics, National University of Singapore, 2 Science Drive 3, Singapore 117542}

\author{Chenliang Su}
\affiliation{SZU-NUS Collaborative Center and International Collaborative Laboratory of 2D Materials for Optoelectronic Science
$\&$ Technology, Engineering Technology Research Center for 2D Materials Information Functional Devices and Systems of Guangdong Province, 
College of Optoelectronic Engineering, Shenzhen University, Shenzhen 518060, China}

\author{Arun Bansil}
\affiliation{Department of Physics, Northeastern University, Boston, Massachusetts 02115, USA}

\author{Hsin~Lin}
\affiliation {Institute of Physics, Academia Sinica, Nankang Taipei 11529, Taiwan}

\date{\today }

\begin{abstract}
The charge density wave (CDW) in 1\textit{T}-TiSe$_2$ harbors a nontrivial symmetry configuration. It is important to understand this underlying symmetry both for gaining a handle on the mechanism of CDW formation and for probing the CDW experimentally. Here, based on first-principles computations within the framework of the density functional theory, we unravel the connection between the symmetries of the normal and CDW states and the electronic structure of 1\textit{T}-TiSe$_2$. Our analysis highlights the key role of irreducible representations (IRs) of the electronic states and the occurrence of band gaps in the system in driving the CDW. By showing how symmetry-related topology can be obtained directly from the electronic structure, our study provides a practical pathway in search of topological CDW insulators. 
\end{abstract}

\maketitle

\section{Introduction}
Transition metal dichalcogenides (TMDCs) have layered crystal structures with the weak van der Waals interaction between the adjacent layers. This feature allows them to be exfoliated in few layers or even monolayers, bringing intensive interest for their appealing electronic and optoelectronic properties \cite{Novoselov2005b, Manzeli2017b, Choi2017, Chen2016, Liu2014g, Velusamy2015}. Owing to the low dimensionality, many materials undergo a charge density wave
(CDW) phase transition \cite{Manzeli2017b,Wilson1974,Whangbo1992}. 1\textit{T}-TiSe$_2$ is a noteworthy examples which shows superconductivity when the CDW is suppressed via doping or pressure \cite{Morosan2006,Barath2008,Kusmartseva2009,Joe2014,Li2016d,Kogar2017a}. The CDW in 1\textit{T}-TiSe$_2$ is commensurate, forming a $2 \times 2 \times 2$ superlattice~\cite{DiSalvo1976,Woo1976}. Simultaneous periodic lattice distortion (PLD)~\cite{Chan1973} was observed from the phonon softening in the $L_{1}^{-}$ mode [$L_{1}^{-}$ is an irreducible representation (IR) at $L$]~\cite{DiSalvo1976,Holt2001,Hildebrand2018}, indicating the symmetry of the CDW. 

The $L_{1}^{-}$ CDW is nontrivial due to its anisotropic character. It is reminiscent of the $d$-density wave in strongly correlated systems where the local rotation symmetry is broken \cite{Li2006a,Davis2013}. Symmetry is important because it bears on the CDW mechanism \cite{Kaneko2018} and determines whether the CDW nodes in the electronic structure influence superconducting $T_{c}$. It also drives directional responses \cite{Boyd2008} and controls the topology of the system. The key role of symmetry, however, has not been hardly explored in the literature~\cite{Monney2009a,Rossnagel2011,VanWezel2010a,Chen2017a}. With this motivation, we will elucidate the symmetry properties of the CDW state in 1\textit{T}-TiSe$_{2}$ and their connections with the electronic structure and the band topology of the system. 

Our analysis starts by considering a system at high temperature, which is described by the bare Hamiltonian $\hat{H}_{0}$ with symmetry group $\mathcal{G}_{0}$. The symmetry elements $R_{i}$ of $\mathcal{G}_{0} $ will of course commute with the Hamiltonian, i.e. $
\left[ \hat{R}_{i},\hat{H}_{0}\right] =0$. When the CDW order develops
at low temperatures, the CDW Hamiltonian becomes $\hat{H}_{\mathrm{CDW}} = \hat{H}_{0} + \hat{\Delta}$, where $\hat{\Delta}$ is
to capture the CDW order. As a result, some symmetries $\hat{R}_{i}$ will be broken when $\left[ \hat{R}_{i},\hat{H}_{\mathrm{CDW}}\right] = \left[ \hat{R}_{i},\hat{\Delta}\right] \neq 0$. Most obvious is the breaking of the translation symmetry $T$ since $\left[ \hat{T},\hat{\Delta} \right] \neq 0$. [For a commensurate CDW the translation symmetry can be restored by enlarging the unit cell: $\hat{T}_{\mathrm{CDW}}=\hat{T}^{M}$ for some integer $M$ such that $\left[ \hat{T}_{\mathrm{CDW}},\hat{\Delta}\right] = 0$.] So, the symmetry group will be reduced to a subgroup $\mathcal{G}$ of $\mathcal{G}_{0}$ which will contain symmetry elements isomorphic to those that are left invariant under the CDW distortion in $\mathcal{G}_{0}$. 

The symmetry of the CDW order parameter (OP)
will be characterized by the symmetry group of the ordering vectors $\mathbf{q}$. We will assume that the OP is a one-dimensional (1D) IR
of the little group $\mathcal{G}_{0}^{\mathbf{q}}$, a subgroup of $\mathcal{G%
}_0$ which contains the point-group symmetries that keep ${\mathbf{q}}$
invariant. By 1D IR we mean that the OP is an eigenstate of point group
symmetry $\hat{R}_{j} \in \mathcal{G}_{0}^{\mathbf{q}}$ with eigenvalue
either $+1$ or $-1$, i.e. $\hat{R}_{j} \hat{\Delta} \hat{R}_{j}^{\dagger} =
\pm \hat{\Delta}$. Note that the broken symmetries in the CDW state also include those associated with eigenvalues of $-1$, which are related to the ordering vectors for the direction of the atomic displacements \cite{VanWezel2010b}.

\section{Symmetry of 1\textit{T}-TiSe$_{2}$}
Pristine 1\textit{T}-TiSe$_{2}$ crystallizes in a hexagonal lattice structure with point group $D_{3d}$ and 
space group $\mathcal{G}_{0}=P\bar{3}m1$.
Its low-temperature phase is the triple-$q$ CDW state with ordering
vectors $\mathbf{Q}_{1}=\frac{1}{2} (\mathbf{a}_1^*+ \mathbf{a}_3^*)$, $%
\mathbf{Q}_{2}=\frac{1}{2} (\mathbf{a}_2^*+ \mathbf{a}_3^*)$, and $\mathbf{Q}%
_{3}=\frac{1}{2} (\mathbf{a}_1^*+\mathbf{a}_2^*+ \mathbf{a}_3^*)$, which
connect $\Gamma $ and the three inequivalent $L$
points ($L_{1,2,3}$), see Fig. \ref{fig:FS}(a). The little group of $L_{2}$
contains point-group elements \{$E,2_{[100]},m_{[100]},\bar{1}$\} as the
group $C_{2h}$, which has four 1D IRs $A_{g},\,B_{g},\,A_{u}$, and $%
B_{u}$ (Table \ref{table:C2h}). With $3_{[001]}$ we obtain conjugate
representations for $L_{2}$ and $L_{3}$. Including translation $\mathbf{T}$
of the Bravais lattice, IRs of the OP can be written as 
\begin{equation}
\begin{split}
\lbrace L_{1}^{+}, L_{2}^{+}, L_{1}^{-}, L_{2}^{-} \rbrace = &\lbrace A_{g},
B_{g}, A_{u}, B_{u} \rbrace \\
& \otimes \left( e^{-i\mathbf{Q}_{1}\cdot \mathbf{T}},e^{-i\mathbf{Q}%
_{2}\cdot \mathbf{T}},e^{-i\mathbf{Q}_{3}\cdot \mathbf{T}}\right).
\end{split}
\label{LIRs}
\end{equation}%
We describe
the OP for a given IR by the vector $\vec{\phi}=(\phi _{1},\phi _{2},\phi
_{3}) \in \mathbb{R}^3$ where the $\vec{\phi}$ accompanying the IR determines the
symmetry group (isotropy group). The energy is better for the maximal isotropy
group \cite{Toledano,Gufan1973} so we will take $\vec{\phi} \propto (1,1,1)$.

Referring to the symmetries of OPs in Table \ref{table:C2h}, we can read
off the space groups for the four CDW states. 
The states with mirror eigenvalue $-1$ do not have a mirror plane but only the glide plane $\lbrace m_{[100]}|0 0 1\rbrace$, where the accompanying $\vec{a}_{3}$ translation produces a complementary $-1$. Similarly states with the parity eigenvalue $-1$ respect the combined symmetry $T_{\mathbf{a}_3} i$ ($=\lbrace \bar{1} |00 1 \rbrace $).
As $T_{\mathbf{a}_{3}} i=T_{\frac{\mathbf{a}_3}{2}} i T_{\frac{\mathbf{a}_3}{%
2}}^{-1}$, there is inversion symmetry with respect the 
center at $\left(0, 0, \frac{1}{2} \right)$. We thus conclude that the space
group for $L_{1}^{+}$ and $L_{2}^{-}$ is $P\bar{3}m1$ (No. 164) and
for $L_{2}^{+}$ and $L_{1}^{-}$ is $P\bar{3}c1$ (No. 165).

\begin{table}[tbp]
\caption{Character table for the point-group symmetry $C_{2h}$ at 
the point $L$. The CDW OP will be one of the IRs $L_{1,2}^{\pm}$. To avoid
confusion, the conduction band at $L$ is labeled with the IR given in parentheses in column 1.
}
\label{table:C2h}
\begin{center}
\setlength{\tabcolsep}{1.5em} 
\begin{tabular}{c|cccc}
\hline
$C_{2h}$ & $E$ & $C_{2}$ & $\sigma _{h}$ & $i$ \\ \hline
$L_{1}^{+}$ ($A_{g}$) & $1$ & $1$ & $1$ & $1$ \\ 
$L_{2}^{+}$ ($B_{g}$) & $1$ & $-1$ & $-1$ & $1$ \\ 
$L_{1}^{-}$ ($A_{u}$) & $1$ & $1$ & $-1$ & $-1$ \\ 
$L_{2}^{-}$ ($B_{u}$) & $1$ & $-1$ & $1$ & $-1$ \\ \hline
\end{tabular}%
\end{center}
\end{table}

\section{Hamiltonian}
The normal state band structure of TiSe$_2$ using the generalized gradient approximation (GGA) without spin-orbit coupling (SOC) is shown in Fig. \ref{fig:bands}(a). There are three bands crossing the Fermi level at $\Gamma$ and one band crossing at $L$. The states at $\Gamma$ consist of the two-fold $E_{g}$ and one-fold $A_{2u}$ representations whereas the state at $L$ belongs to the $A_g$ representation. The mean field $k\cdot p$ Hamiltonian for the CDW can be expressed as 
\begin{equation} \label{HCDW}
\hat{H}_{\mathrm{CDW}}=\sideset{}{'}\sum_{\mathbf{k}}\psi _{\mathbf{k}%
}^{\dagger }\mathcal{H}_{\mathrm{CDW}}(\mathbf{k})\psi _{\mathbf{k}},
\end{equation}%
where the basis $\psi _{\mathbf{k}}^{\dagger }=(\psi _{\Gamma \mathbf{k}%
}^{\dagger },~\psi _{L\mathbf{k}}^{\dagger },\psi _{A\mathbf{k}}^{\dagger
},~\psi _{M\mathbf{k}}^{\dagger })$ is taken and 
\begin{equation}
\mathcal{H}_{\mathrm{CDW}}(\mathbf{k})=\left( 
\begin{array}{cc}
\mathcal{H}_{\Gamma L} & \mathcal{V} \\ 
\mathcal{V}^{\dagger } & \mathcal{H}_{AM}%
\end{array}%
\right) (\mathbf{k}).
\end{equation}%
The prime on the summation in Eq.~(\ref{HCDW}) indicates that only low-energy or $\mathbf{k}$ states in the folded Brillouin zone (BZ) are considered (spin index is omitted for brevity). The states at the $\Gamma $ and 
$L$'s points as well as at the $A$ and $M$ points are coupled by the CDW gaps. Also, because $q$ vectors connect the $\Gamma$ and $A$ points, these four points can be written together. The states at 
$\Gamma $ and $L$'s can be described by $\mathcal{H}_{\Gamma L}$:   
\begin{equation}
\mathcal{H}_{\Gamma L}(\mathbf{k})=\left( 
\begin{array}{cc}
\mathcal{H}_{\Gamma } & \Delta _{\Gamma L} \\ 
\Delta _{\Gamma L}^{\dagger } & \mathcal{H}_{L}%
\end{array}%
\right) (\mathbf{k}),
\end{equation}%
where $\mathcal{H}_{\Gamma }$ and $\mathcal{H}_{L}$ account for the
normal-state band structure around $\Gamma $ and $L$'s, respectively. 
In $%
\psi _{\Gamma \mathbf{k}}^{\dagger }=(\psi _{\Gamma _{1}\mathbf{k}}^{\dagger
},~\psi _{\Gamma _{2}\mathbf{k}}^{\dagger },~\psi _{\Gamma _{3}\mathbf{k}%
}^{\dagger })$, the former two operators stand for the $E_{g}$ states and the
last one is for $A_{2u}$. In $\psi _{L\mathbf{k}}=(\psi _{L_{1}\mathbf{k}%
}^{\dagger },~\psi _{L_{2}\mathbf{k}}^{\dagger },~\psi _{L_{3}\mathbf{k}%
}^{\dagger })$, the three bands at the three $L$'s are included with $\mathbf{k}$
being relative to the corresponding $L$'s. The CDW gap $\Delta _{\Gamma L}$
is a $3\times 3$ matrix, where the matrix element $\Delta _{ij}$ refers to the
CDW potential between $\psi _{\Gamma _{i}}^{\dagger }$ and $\psi _{L_{j}}$.
We will mainly discuss $\mathcal{H}_{\Gamma L}$ and
$\mathcal{H}_{AM}$. The off-diagonal element $\mathcal{V}$, which is induced by the second secondary OP, is expected to be weak.

The Hamiltonians can be determined using the 
symmetry constraints as follows:
\begin{eqnarray}
\mathcal{C}_{3\mathrm{CDW}}\mathcal{H}_{\mathrm{CDW}}(\mathbf{k})\mathcal{C}%
_{3\mathrm{CDW}}^{-1} &=&\mathcal{H}_{\mathrm{CDW}}(\mathfrak{R}\mathbf{k}),
\label{symmetry1} \\
\mathcal{I}_{\mathrm{CDW}}\mathcal{H}_{\mathrm{CDW}}(\mathbf{k})\mathcal{I}_{%
\mathrm{CDW}}^{-1} &=&\mathcal{H}_{\mathrm{CDW}}(\mathfrak{-}\mathbf{k}),
\label{symmetry2} \\
\mathcal{M}_{\mathrm{CDW}}\mathcal{H}_{\mathrm{CDW}}(\mathbf{k})\mathcal{M}_{%
\mathrm{CDW}}^{-1} &=&\mathcal{H}_{\mathrm{CDW}}(\mathfrak{M}\mathbf{k}),
\label{symmetry3}
\end{eqnarray}%
where $\mathfrak{R}\mathbf{k}$, $\mathfrak{-}\mathbf{k}$, and $\mathfrak{M}%
\mathbf{k}$ are momenta under $3_{[001]}$, $\bar{1}$, and $m_{[100]}$,
respectively. We define the $[100]$ direction to be $x$ and $[001]$
to be $z$. The relevant symmetry operators are 
\begin{equation}
\mathcal{O}_{\mathrm{CDW}}=\left( 
\begin{array}{cc}
\mathcal{O}_{\mathrm{CDW,}\Gamma L} &  \\ 
& \mathcal{O}_{\mathrm{CDW,}AM}%
\end{array}%
\right) ,
\end{equation}%
with%
\begin{equation}
\mathcal{O}_{\mathrm{CDW,}\Gamma L}=\left( 
\begin{array}{cc}
\mathcal{O}_{\Gamma} &  \\ 
& \eta _{\mathcal{O}}\mathcal{O}_{L}%
\end{array}%
\right) 
\end{equation}
and
\begin{equation}
\mathcal{O}_{\mathrm{CDW,}AM}=\left( 
\begin{array}{cc}
\eta _{\mathcal{O}}^3\mathcal{O}_{A} &  \\ 
& \eta _{\mathcal{O}}^2\mathcal{O}_{M}%
\end{array}%
\right) 
\end{equation}%
where $\mathcal{O}=\{\mathcal{C}_{3},\mathcal{I},\mathcal{M}\}$ and $\eta _{\mathcal{O}}$ is the eigenvalue of the OP for operation $\mathcal{O}$. 
Note that the eigenvalue in front of $\mathcal{O}_A$ takes $\eta _{\mathcal{O}}^3$ because $\Gamma$ is connected to $A$ via the sum of three ordering vectors $\mathbf{Q}_i$. But, the eigenvalue at $\mathcal{O}_M$ takes $\eta _{\mathcal{O}}^2$ because a sum of two $\mathbf{Q}_i$ connects $\Gamma$ and $M$. As $%
\mathcal{O}_{\mathrm{CDW}}$ is block-diagonal, Eqs. (\ref{symmetry1})-(\ref%
{symmetry3}) determine the form of $\mathcal{H}_{\Gamma }$, $\mathcal{H}%
_{L}$, and $\Delta_{\Gamma L} $, see Appendix \ref{H0_app} for details. 
$\eta_{\mathcal{O}}$ reflects the symmetries of the OP via the CDW gap matrix as follows 
\begin{eqnarray}
\mathcal{C}_{3,\Gamma }\Delta _{\Gamma L}(\mathbf{k})\mathcal{C}_{3,L}^{-1}
&=&\eta _{\mathcal{C}_{3}}\Delta _{\Gamma L}(\mathfrak{R}\mathbf{k}),
\label{gap_s1} \\
\mathcal{I}_{\Gamma }\Delta _{\Gamma L}(\mathbf{k})\mathcal{I}_{L}^{-1}
&=&\eta _{\mathcal{I}}\Delta _{\Gamma L}(\mathfrak{-}\mathbf{k}),
\label{gap_s2} \\
\mathcal{M}_{\Gamma }\Delta _{\Gamma L}(\mathbf{k})\mathcal{M}_{L}^{-1}
&=&\eta _{\mathcal{M}}\Delta _{\Gamma L}(\mathfrak{M}\mathbf{k}),  \label{gap_s3}
\end{eqnarray}%
where $\eta _{\mathcal{I}}$, $\eta _{\mathcal{M}}$ ($=\pm 1$) and $\eta _{%
\mathcal{C}_{3}}=1$. The four IRs $L_{1}^{+},~L_{2}^{+},~L_{1}^{-},~L_{2}^{-}$ are obtained
by taking $(\eta _{\mathcal{I}},\eta _{\mathcal{M}})=(1,1),~(1,-1),~(-1,-1)$%
, and $(-1,1)$, respectively (listed also in Table \ref{table:C2h}).

Although the $L_1^{-}$ CDW OP has an $f$-wave symmetry \cite{CastroNeto2001}, the CDW gap functions may be of different symmetry depending on IRs of the composite bands as we will show below. It can be shown that the gap functions for the $L_1^{-}$ CDW state to the lowest
order of $k$ are:
\begin{align}
\begin{split}
\Delta _{12}(\mathbf{k})=\lambda _{11}k_{x},~\Delta _{22}(%
\mathbf{k})=\lambda _{21}k_{y}+\lambda _{21}^{\prime }k_{z}, \\
\Delta _{32}(\mathbf{k})=(\lambda _{31}k_{y}+\lambda _{31}^{\prime
}k_{z})k_{x},
\end{split} \label{gapL1m}
\end{align}%
where $\lambda $, $\lambda ^{\prime }$'s are taken to be real. The other terms can be obtained using
the rotation symmetry, see Appendix \ref{CDWgap_app} for details.
The gap functions give the couplings among the valence bands at $\Gamma$ and the folded conduction bands from the $L$ points. 
Assuming that the valence and conduction bands in the normal state overlap, in the small gap limit, the gap functions in Eq. (\ref{gapL1m}) will manifest whether bands cross or anticross according to their mirror or twofold rotation eigenvalues. For the large gap case, where multi-band hybridizations will be involved, this two-band picture will fail, and we will resort to topological protections. 

Figure \ref{fig:bands}(b) presents the band structure of the CDW state in the presence of PLD. We can see that the CDW gap size ($\sim 0.1$ eV) is not small and the band structure is complex as a result of many band foldings. Now there are two CDW nodes (band crossings between the conduction and valence bands), one on the $\overline{\Gamma K}$ line and the other along the $\overline{\Gamma M}$ line. Taking into account the rotation and time-reversal symmetries, there are 12 nodes on the $k_z=0$ plane. A full BZ exploration of the band structure shows that these nodes persist away from the $k_z=0$ plane and yield nodal lines in the BZ, see Fig. \ref{fig:bands}(c). Band structures for different $k_z$ values are given in Appendix \ref{bands_kz}.

\section{Ginzburg-Landau theory}
We turn now to discuss the Ginzburg-Landau theory for the $L_{1}^{-}$ CDW state. 
Images of the primary OP ($L_{1}^{-}$ mode) $\vec{\phi} =
\left(\phi_1,\phi_2,\phi_3\right)$ under various symmetry operations are: 
\begin{equation}
\begin{split}
3_{[001]} : {}& ~ \left(\phi_1,\phi_2,\phi_3\right) \rightarrow
\left(\phi_2,\phi_3,\phi_1\right), \\
m_{[100]}:{}& ~ \left(\phi_1,\phi_2,\phi_3\right) \rightarrow
\left(-\phi_3,-\phi_2,-\phi_1\right), \\
\bar{I} :{}& ~ \left(\phi_1,\phi_2,\phi_3\right) \rightarrow
\left(-\phi_1,-\phi_2,-\phi_3\right), \\
T_{\mathbf{a}_1} :{}& ~ \left(\phi_1,\phi_2,\phi_3\right) \rightarrow
\left(-\phi_1,\phi_2,-\phi_3\right), \\
T_{\mathbf{a}_2} :{}& ~ \left(\phi_1,\phi_2,\phi_3\right) \rightarrow
\left(\phi_1,-\phi_2,-\phi_3\right), \\
T_{\mathbf{a}_3} :{}& ~ \left(\phi_1,\phi_2,\phi_3\right) \rightarrow
\left(-\phi_1,-\phi_2,-\phi_3\right), \\
\end{split}
\label{sym_primaryOP}
\end{equation}
so that the free-energy density which satisfies symmetries of the pristine state to
the fourth order becomes:
\begin{equation}
F_{\mathrm{primary}} =\alpha \vec{\phi}^2 + \beta_{1} \vec{\phi}^4 +
\beta_{2} \left( \phi_1^4 + \phi_2^4 + \phi_3^4 \right).  \label{Fprimary}
\end{equation}
Here $\alpha<0$ when $T<T_c$ for nonzero $\vec{\phi}$ and $\beta>0$ for
stability. Therefore, we seek a secondary OP that is linearly coupled to the primary OP $\vec{\phi}$, i.e. we look for coincidence of the primary and secondary
OPs (Appendix \ref{GL_app}). Symmetry requirements suggest that the
secondary OP $\vec{\zeta}$ will also have three components with the  coupling energy
\begin{equation}
F_{\mathrm{coupling}} = \lambda \left( \phi_1 \phi_2 \zeta_3 + \phi_2 \phi_3
\zeta_1 + \phi_3 \phi_1 \zeta_2\right),
\end{equation}
where $\vec{\zeta}$ follows the rules in Eq. (\ref{sym_primaryOP})
except that $m_{[100]}$, $\bar{I}$, and $T_{\mathbf{a}_3}$ are without the minus signs.
The symmetry conditions dictate that the components of the secondary OP
take ordering vectors $\mathbf{Q}^{\prime }_{1}=\frac{1}{2} \mathbf{a}_1^*$, 
$\mathbf{Q}^{\prime }_{2}=\frac{1}{2} \mathbf{a}_2^*$, and $\mathbf{Q}%
^{\prime }_{3}=\frac{1}{2}(\mathbf{a}_1^*+\mathbf{a}_2^*)$ and that they belong
to IR $M_{1}^{+}$. In addition to the terms similar to Eq. (\ref{Fprimary}), the
free energy for the secondary OP can contain the cubic
term $\zeta_1 \zeta_2 \zeta_3$. Besides, coexistence of $L_{1}^{-}$ and 
$M_{1}^{+}$ OPs will induce a third (second secondary) OP $\varphi$ in higher order as 
\begin{equation}
F_{\mathrm{coupling}}^{\prime }= \lambda^{\prime }\varphi \left( \phi_1
\zeta_1 + \phi_2 \zeta_2+ \phi_3 \zeta_3\right),
\end{equation}
where $\varphi$, which we call IR $A_{2}^{-}$, is parity-odd, mirror-odd and belongs to the 
ordering vector $\frac{1}{2} \mathbf{a}_3^* $. Close to $T_c$, $\zeta$ is proportional to $\phi$ square and thus $\psi$ is proportional to $\phi$ cube, so that the symmetry characters of $M_1^{+}$ ($=L_1^{-} \otimes L_1^{-}$) are the same as $ L_1^{+}$ and those of $A_{2}^{-}$ are identical to $L_1^{-}$. Incidentally, the coincident $M_{1}^{+}$ OP will also be present in the other three CDW states, accompanied by the corresponding third OPs.

\section{IRs of bands}
Although the gap functions in Eq. (\ref{gapL1m}) all vanish at $\mathbf{k=0}$, the folded bands still result in threefold degeneracy. But, this threefold degeneracy should not occur as it is not robust in the $D_{3d}$ group. This can be understood, however, because the concurrent $M_1^{+}$ mode produces couplings among the bands at $L$ points (also among the $M$ points): the triplet from $L$'s then splits into a singlet $A_{1u}$ state and a doublet $E_u$ state without breaking symmetry \cite{Kidd2002}. We emphasize that the parity of the folded bands changes from even to odd due to the CDW OP. In general, the IR of a folded band is the product of IRs of the unfolded band and the OP, written as:
\begin{equation}
\Gamma^{\mathbf{k}}_{\mathrm{folded}} = \Gamma^{\mathbf{k-Q}}_{\mathrm{unfolded}} \otimes \Gamma^{\mathbf{Q}}_{\mathrm{CDW}}. \label{irrep_band}
\end{equation} 
This is our key finding, which is especially useful for investigating topology. By contrast, folded bands from $M$'s to $\Gamma$, through the secondary $M_1^{+}$ OP, split into an $A_g$ and two $E_g$ states. As for the bands from $A$, their parities change sign because of the $A_{2}^{-}$ mode. These arguments are well corroborated by our independent first-principles band structure results, see Fig. \ref{fig:bands}(b).

\begin{figure}[t]
\centering
\includegraphics[width=0.45\textwidth]{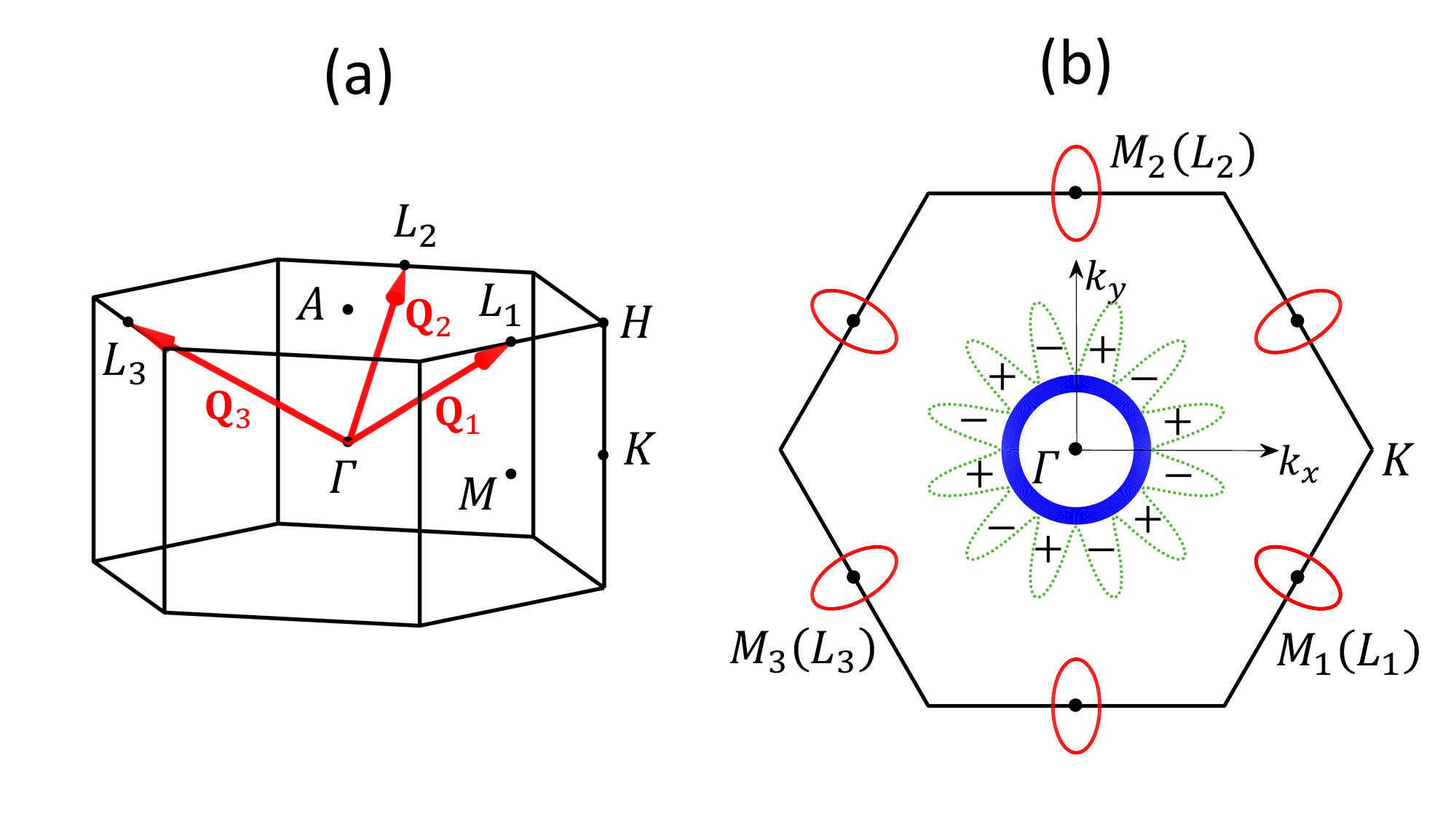}
\caption{(a) Bulk BZ and the CDW ordering vectors $\mathbf{Q}_{1,2,3}$ of 1$T$-TiSe$_2$. (b) Schematic Fermi surface (FS) of the pristine 1$T$-TiSe$_2$ in the $k_z=0$ plane. Three hole-pockets at $\Gamma$ are marked by the thick blue circle. Electron pockets at the $L$ points are shown by red ellipses. The green dotted lines around $\Gamma$ show a $i$-wave CDW gap function, which supports 12 line nodes on the FS of the CDW state.}
\label{fig:FS}
\end{figure}

\begin{figure*}[t]
\centering
\includegraphics[width=1\textwidth]{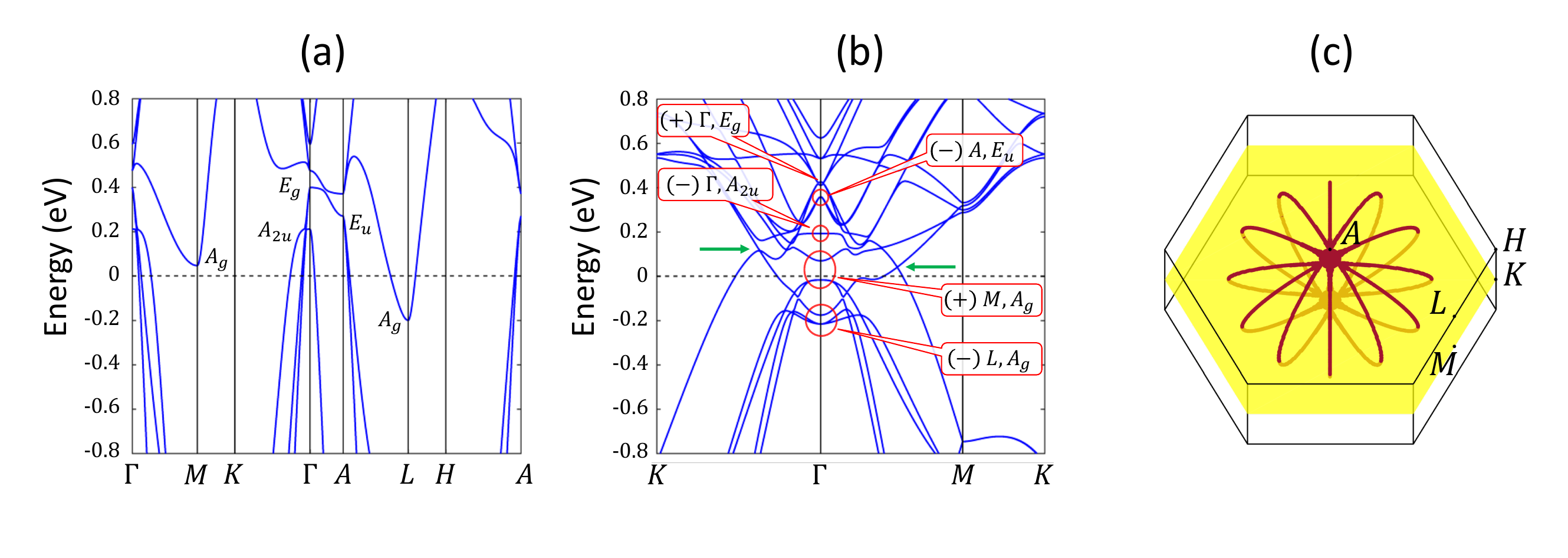}
\caption{Band structures without spin-orbit coupling for (a) the normal state and (b) the $L_{1}^{-}$ CDW state. Energies are given relative to the Fermi energy. In (a), the IRs for relevant states at the symmetry points are listed. In (b), the parity eigenvalues of states at $\Gamma$ are given in parentheses along with the original $k$ points and IRs. Green arrows point to the CDW nodes. (c) Illustration of the 12 nodal lines piercing through the $k_z=0$ plane (in yellow) in the folded BZ. The nodal lines were determined from the zeros of the CDW gap function on the FS based on $\Delta_{32}(\mathbf{k})$ in Eq. (\ref{gapL1m}). }
\label{fig:bands}
\end{figure*}

Note that the gap function of two composite bands has the symmetry of the direct product of the IRs of the two bands. For instance, if the closest conduction band is IR $A_{1u}$ and the valence band is IR $A_{2u}$, the associated gap function will be IR $A_{2g}$ ($=A_{1u} \otimes A_{2u}$) that is also known as the $i$ wave outlining a $\sin(6 \phi)$ profile, Fig. \ref{fig:FS}(b).

\section{Topology}
We recall at the outset \cite{Kim2015d} that in the presence of inversion and time-reversal symmetries, SOC-free systems can be classified into semimetals with even or odd number of bulk line nodes, which can be labeled with a set of $\mathbb{Z}_{2}$ invariants obtained from the product of parity eigenvalues of filled bands at the four parity-invariant momenta in a BZ plane. The line nodes are gapped by the SOC to yield weak topological insulators \cite{Fu2007b,Fu2007c}.

In the CDW state, the parity-product at $M$, $L$, and $A$ points in the folded BZ will be trivial, so that only the $\Gamma$ point remains relevant. The reason is that the original and folded bands hybridize into symmetric and antisymmetric combinations resulting in a parity-product of $-1$, and even number of such pairs will yield a net product of $+1$. In the case of no band inversion, the parity-product at $\Gamma$ of the CDW state is related to the strong topological invariant of the pristine state. Therefore, the strong topological invariant of the CDW state will be determined by the number (mod 2) of band inversions which involve parity switching beyond the topology of the normal state. 

Our DFT calculations show that the pristine and the $L_1^{-}$ CDW states are both topologically trivial. Figure \ref{fig:bands}(b) shows that the three bands at $\Gamma$  ($E_{u}$ and $A_{2u}$ bands) hybridize with three pristine bands at $L$ points without changing parity yielding a net parity-product of $+1$. Also, the two bands at $A$ ($E_{u}$ symmetry) hybridize with two of three bands at the $M$ points and maintain the parity-product of $+1$. An even number of nodal rings is thus expected to pierce through the $k_z=0$ plane as is seen to be the case in Fig. \ref{fig:bands}(c). In this connection, we have further analyzed the band structure of the $L_1^{+}$ CDW state (not shown for brevity) to show that it is a topologically nontrivial semimetal consistent with Eq. (\ref{irrep_band}).

\section{Conclusion and discussion}
We have presented an in-depth analysis of the symmetries as well as the topology of the electronic structure of bulk 1$T$-TiSe$_2$ CDW. Our first-principles calculations show that the CDW state hosts a nodal band structure in which the nodes are protected by symmetry and topology resembling that of the Dirac nodes in the spin-density-wave phase of iron pnictides \cite{Ran2009, Richard2010}. The existing topological theory in this connection only considers spin-density-wave states~\cite{Mong2010, Fang2013} in terms of $\mathbb{Z}_2$ classification of three-dimensional insulators, but questions of symmetry properties of the density waves and their connection to the normal states have not been addressed. We resolve these questions by successfully connecting the symmetry and topology of the electronic IRs of the normal and CDW phases. 

We emphasize that in our theory when both the time-reversal and inversion symmetries are preserved, inclusion of the SOC gaps band crossings for out-of-plane $n$-fold ($n>2$) rotation axis to protect the Dirac nodes, but it does not change the parity of the bands. Although the CDW OP might change the IR and parity of the folded band, it does not always produce band inversion. We discuss the application of our theory to the band structure based on the GGA density functional, which yields a semimetallic normal state with an energy overlap (indirect negative band gap) between the conduction and valence bands. On the experimental side, there has been a longstanding debate whether the normal state of 1$T$-TiSe$_2$ is a semimetal~\cite{Anderson1985, Li2007, Cercellier2007} or a semiconductor~\cite{Kidd2002, Rasch2008}. Our study however is not concerned with such details of the band structure. Our results are intended to be generic in nature and are not sensitive to the density functional used. The question of sensitivity of the band structure of 1\textit{T}-TiSe$_2$ to exchange-correlation functional, Hubbard $U$, and van der Waals interactions has been explored extensively in the literature~\cite{Bianco2015, Hellgren2017,Hellgren2021}. Correlations are generally found to reduce the band overlap in better accord with experimental results. This is also the case in our first-principles band structure of 1\textit{T}-TiSe$_2$ obtained with the modified Becke-Johnson (mBJ) meta-GGA density functional, which correctly captures the bandgap correction and reproduces an insulating electronic state, see Appendix~\ref{bands_pressure} for details. Our theory is thus a ``weak-pairing theory" in which the Fermi energy is larger than the CDW gap. For these reasons, it is natural for us to base our analysis on the GGA band structure.

\begin{figure}[t]
\centering
\includegraphics[width=0.45\textwidth]{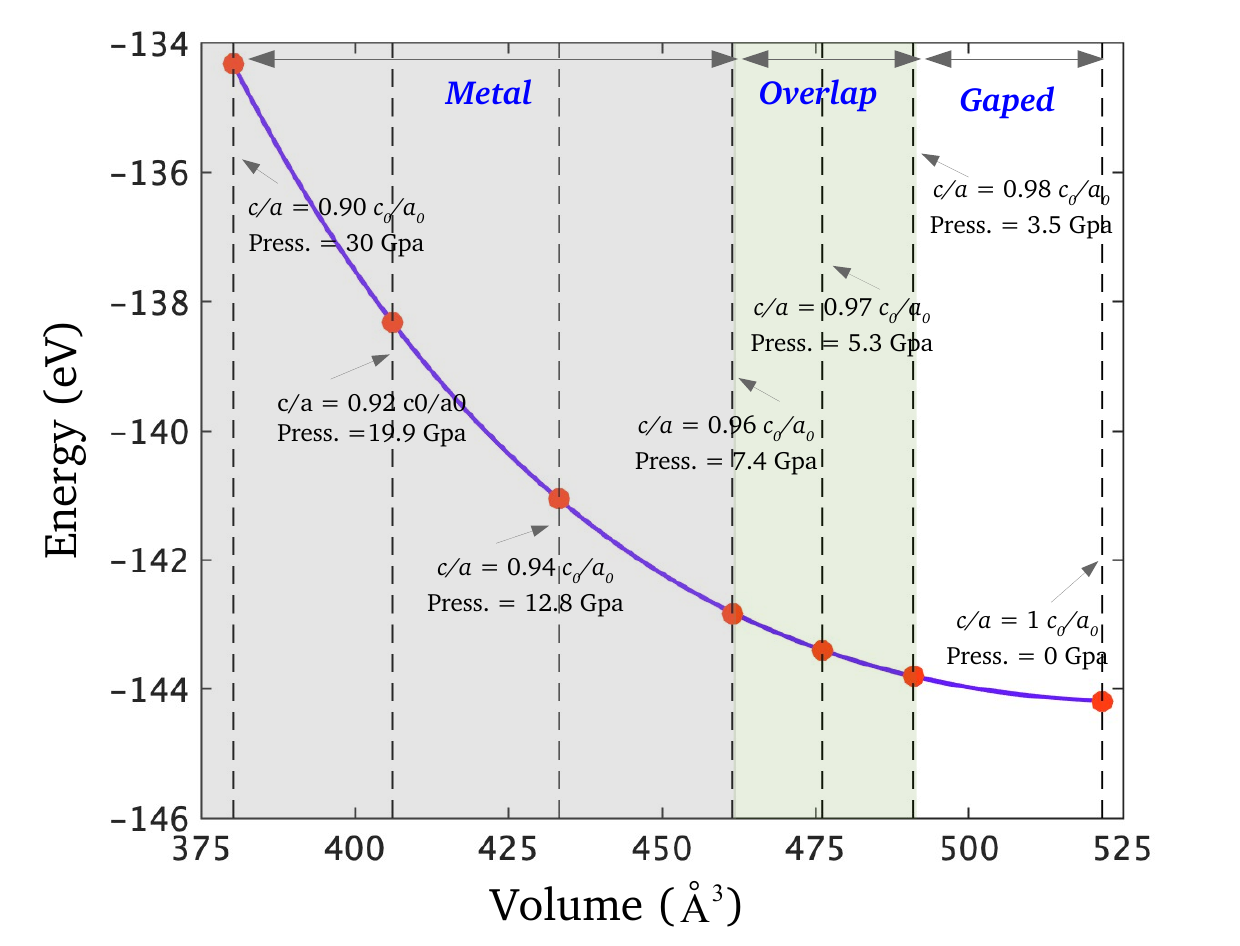}
\caption{Energy as a function of volume (hydrostatic pressure) of $2\times2\times2$ bulk 1$T$-TiSe$_2$ obtained using the modified Becke-Johnson (mBJ) meta-GGA density functional without SOC. The change in lattice parameters and associated hydrostatic pressure are marked. Color regions identify the CDW gapped (white), CDW semimetal (green), and normal metal (gray) states. The system evolves from a large gap CDW insulator to a CDW semimetal for $c/a=0.97 c_0/a_0$) and a non-CDW metallic state for $c/a=0.96 c_0/a_0$. $a_{0}$ and $c_{0}$ ($a$ and $c$) define the lattice constants at zero (finite) pressure.  }
\label{fig:pressure}
\end{figure}

Figure~\ref{fig:pressure} shows the pressure effect on the electronic structure of bulk 1\textit{T}-TiSe$_2$. The energy overlap is seen to be retained under moderate pressures as is the CDW order. This suggests that a topological phase transition in the CDW state can be achieved by applying hydrostatic pressure. By showing how symmetry-related topology can be obtained directly from the electronic structure, our study provides a guide in search of topological CDW phases. Our analysis can be generalized straightforwardly to consider other spontaneous symmetry-breaking phases.

\begin{acknowledgments}
S.M.H is supported by the Ministry of Science and Technology (MoST) in Taiwan under grant No. 105-2112-M-110-014-MY3 and also by the NCTS of Taiwan. He also acknowledges support from Academia Sinica's Short-term Visiting Program for Domestic Scholars. The work at TIFR Mumbai is supported by the Department of Atomic Energy of the Government of India under Project No. 12-R$\&$D-TFR- 5.10-0100. The work at Northeastern University was supported by the US Department of Energy (DOE), Office of Science, Basic Energy Sciences Grant No. DE-SC0019275 (materials discovery for QIS applications) and benefited from Northeastern University's Advanced Scientific Computation Center and the National Energy Research Scientific Computing Center through DOE Grant No. DE-AC02-05CH11231.
\end{acknowledgments}

\appendix
\section{Hamiltonian without spin-orbit coupling}
Here we discuss the construction of our mean-field $2\times 2 \times 2$ charge density wave (CDW) Hamiltonian without spin-orbit coupling for bulk 1\textit{T}-TiSe$_2$. It is written as
\begin{equation}
\hat{H}_{\mathrm{CDW}}=\sideset{}{'}\sum_{\mathbf{k}}\psi _{\mathbf{k}%
}^{\dagger }\mathcal{H}_{\mathrm{CDW}}(\mathbf{k})\psi _{\mathbf{k}},
\end{equation}%
where the prime indicates that $\mathbf{k}$ runs over states near the Fermi
surfaces and 
\begin{equation}
\mathcal{H}_{\mathrm{CDW}}(\mathbf{k})=\left( 
\begin{array}{cc}
\mathcal{H}_{\Gamma L} & \mathcal{V} \\ 
\mathcal{V}^{\dagger } & \mathcal{H}_{AM}%
\end{array}%
\right) (\mathbf{k})
\end{equation}%
in the basis $\psi _{\mathbf{k}}^{\dagger }=(\psi _{\Gamma \mathbf{k}%
}^{\dagger },~\psi _{L\mathbf{k}}^{\dagger },\psi _{A\mathbf{k}}^{\dagger
},~\psi _{M\mathbf{k}}^{\dagger })$. 
Based on our first-principles calculations, the normal-state band structure near the Fermi level consists of three valence states at the $\Gamma $ point, two valence states at the $A$ point, and a conduction band whose minimum locates at each $L$ point. The $L$-point conduction band adiabatically evolves into bands at the $M$ point, see Fig. 2(a) of the main text. The two valence bands at $\Gamma$ are comprised of $E_{g}$ IR whereas the third band belongs to $A_{2u}$ (Table \ref{table:D3d}). The bands at $A$ have the IR $E_{u}$ whereas at the $L$ and $M$, they have $A_{g}$ IR (Table \ref{table:C2h}).
For $\psi _{\Gamma \mathbf{k}%
}^{\dagger }$, which contains three creation operators for the hole pockets
around $\Gamma $, $\psi _{\Gamma \mathbf{k}}^{\dagger }=(\psi _{\Gamma _{1}%
\mathbf{k}}^{\dagger },~\psi _{\Gamma _{2}\mathbf{k}}^{\dagger },~\psi
_{\Gamma _{3}\mathbf{k}}^{\dagger })$, we assign $E_{g}$ to the first
two operators and $A_{2u}$ to the last operator. For $\psi _{A \mathbf{k}%
}^{\dagger }$, we denote $\psi _{A \mathbf{k}}^{\dagger }=(\psi _{A_1 \mathbf{k}}^{\dagger },~\psi _{A_2 \mathbf{k}}^{\dagger })$. As for $\psi _{L\mathbf{k}}$%
, it contains three creation operators for the electron pockets around the three $L$ points, $\psi _{L\mathbf{k}}^{\dagger}=(\psi _{L_{1}\mathbf{k}}^{\dagger},~\psi _{L_{2}\mathbf{k}}^{\dagger },~\psi _{L_{3}\mathbf{k}}^{\dagger })$. 
Replacing $L$ by $M$, we obtain $\psi _{M\mathbf{k}}^{\dagger}$.

We will mainly consider the CDW potentials among the bands at $\Gamma$ and $L$ points. Other potentials can be constructed along similar lines. We first examine
\begin{equation}
\mathcal{H}_{\Gamma L}(\mathbf{k})=\left( 
\begin{array}{cc}
\mathcal{H}_{\Gamma } & \Delta _{\Gamma L} \\ 
\Delta _{\Gamma L}^{\dagger } & \mathcal{H}_{L}%
\end{array}%
\right) (\mathbf{k}),
\end{equation}%
where $\mathcal{H}_{\Gamma }$ and $\mathcal{H}_{L}$ account for the
normal-state band structure around $\Gamma $ and $L$'s, respectively. 
In these expression, $\mathcal{H}_{\Gamma }$, $\mathcal{H}_{L}$ and $\Delta _{\Gamma L}$ are all $3\times 3$ matrices written as 
\begin{eqnarray}
\mathcal{H}_{\Gamma }(\mathbf{k}) &=&\left( 
\begin{array}{ccc}
h_{11} & h_{12} & h_{13} \\ 
h_{12}^{\ast } & h_{22} & h_{23} \\ 
h_{13}^{\ast } & h_{23}^{\ast } & h_{33}%
\end{array}%
\right) (\mathbf{k}), \\
\mathcal{H}_{L}(\mathbf{k}) &=&\left( 
\begin{array}{ccc}
\xi _{1} & 0 & 0 \\ 
0 & \xi _{2} & 0 \\ 
0 & 0 & \xi _{3}%
\end{array}%
\right) (\mathbf{k}), \\
\Delta _{\Gamma L}(\mathbf{k}) &=&\left( 
\begin{array}{ccc}
\Delta _{11} & \Delta _{12} & \Delta _{13} \\ 
\Delta _{21} & \Delta _{22} & \Delta _{23} \\ 
\Delta _{31} & \Delta _{32} & \Delta _{33}%
\end{array}%
\right) (\mathbf{k}).
\end{eqnarray}%
$\Delta _{ij}(\mathbf{k})$ will be regarded as the CDW potential between $%
\Gamma _{i}$ and $L_{j}$. The Hamiltonians can then be written down according to
the symmetry constraints as: 
\begin{eqnarray}
\mathcal{C}_{3}\mathcal{H}_{\mathrm{CDW}}(\mathbf{k})\mathcal{C}%
_{3}^{-1} &=&\mathcal{H}_{\mathrm{CDW}}(\mathfrak{R}\mathbf{k}),
\label{symmetry1app} \\
\mathcal{IH}_{\mathrm{CDW}}(\mathbf{k})\mathcal{I}^{-1} &=&\mathcal{H}_{%
\mathrm{CDW}}(\mathfrak{-}\mathbf{k}),  \label{symmetry2app} \\
\mathcal{M}\mathcal{H}_{\mathrm{CDW}}(\mathbf{k})\mathcal{M}^{-1} &=&\mathcal{H}_{\mathrm{CDW}}(\mathfrak{M}\mathbf{k}),
\label{symmetry3app}
\end{eqnarray}%
where $\mathfrak{R}\mathbf{k}$, $\mathfrak{-}\mathbf{k}$, and $\mathfrak{M}%
\mathbf{k}$ are momenta under (counterclockwise) $3_{[001]}$, $\bar{1}$, and 
$m_{[100]}$, respectively. We define the $[100]$ direction to be $x$ and $[001]$ to be $z$. The relevant symmetry operators would be $k$%
-dependent and are block-diagonal as 
\begin{widetext}
\begin{equation}
\mathcal{C}_{3}=\left( 
\begin{array}{cccc}
\mathcal{C}_{3,\Gamma } & ~ &~ & ~ \\ 
~  & \mathcal{C}_{3,L} & ~ & ~  \\
  ~ & ~ & \mathcal{C}_{3,A}  & ~ \\
  ~ & ~ & ~ & \mathcal{C}_{3,M} 
\end{array}%
\right) ,~\mathcal{I}=\left( 
\begin{array}{cccc}
\mathcal{I}_{\Gamma } & ~ & ~ & ~\\ 
~ & \mathcal{I}_{L} & ~ & ~ \\
~ & ~ & \mathcal{I}_{A}  & ~ \\
~ & ~ & ~ & \mathcal{I}_{M} 
\end{array}%
\right) ,~\mathcal{M}=\left( 
\begin{array}{cccc}
\mathcal{M}_{\Gamma } & ~ & ~ & ~\\ 
~ & \mathcal{M}_{L} & ~ & ~ \\
 ~ & ~ & \mathcal{M}_{A}  & ~ \\
  ~ & ~ & ~ & \mathcal{M}_{M} 
\end{array}%
\right) .
\end{equation}%
We choose 
\begin{eqnarray}
\mathcal{C}_{3,\Gamma } &=&\left( 
\begin{array}{ccc}
-\frac{1}{2} & \frac{\sqrt{3}}{2} & 0 \\ 
-\frac{\sqrt{3}}{2} & -\frac{1}{2} & 0 \\ 
0 & 0 & 1%
\end{array}%
\right) ,~
\mathcal{C}_{3,A} = \left( 
\begin{array}{cc}
-\frac{1}{2} & \frac{\sqrt{3}}{2}  \\ 
-\frac{\sqrt{3}}{2} & -\frac{1}{2} 
\end{array}%
\right) ,~
\mathcal{C}_{3,L}=\mathcal{C}_{3,M}=\left( 
\begin{array}{ccc}
0 & 0 & 1 \\ 
1 & 0 & 0 \\ 
0 & 1 & 0%
\end{array}%
\right) , \\
\mathcal{I}_{\Gamma } &=&\left( 
\begin{array}{ccc}
1 & 0 & 0 \\ 
0 & 1 & 0 \\ 
0 & 0 & -1%
\end{array}%
\right) ,~
\mathcal{I}_{A} =\left( 
\begin{array}{cc}
-1 & 0 \\ 
0 & -1
\end{array}%
\right) ,~
\mathcal{I}_{L}=\mathcal{I}_{M}=\left( 
\begin{array}{ccc}
1 & 0 & 0 \\ 
0 & 1 & 0 \\ 
0 & 0 & 1%
\end{array}%
\right) , \\
\mathcal{M}_{\Gamma } &=&\left( 
\begin{array}{ccc}
1 & 0 & 0 \\ 
0 & -1 & 0 \\ 
0 & 0 & 1%
\end{array}%
\right) ,~
\mathcal{M}_{A } = \left( 
\begin{array}{cc}
1 & 0 \\ 
0 & -1
\end{array}%
\right) ,~
\mathcal{M}_{L}=\mathcal{M}_{M}=\left( 
\begin{array}{ccc}
0 & 0 & 1 \\ 
0 & 1 & 0 \\ 
1 & 0 & 0%
\end{array}%
\right) .
\end{eqnarray}%
\end{widetext}
Note that here we adopt a representation in which $\mathcal{I}_{L}$, $\mathcal{I}_{L}$, $\mathcal{M}_{\Gamma }$ and $\mathcal{M}_{A}$ matrices are diagonal, where the diagonal elements are the eigenvalues of the corresponding symmetries at various symmetry points. In this basis, the $\mathcal{C}_{3,\Gamma }$ and $\mathcal{C}_{3,A }$ matrices are non-diagonal
for the $E_{g}$ and $E_{u}$ states. As for the $L$ and $M$ points, they involve large momentum
transfers. For instance, we set $\left\vert \psi _{L_{1}\mathbf{k}%
}\right\rangle \mapsto \left\vert \psi _{L_{2}\mathfrak{R}\mathbf{k}%
}\right\rangle $ under $3_{[001]}$; while under $m_{[100]}$, which leaves $L_{2}$
invariant and interchanges $L_{1}$ and $L_{3}$, $\left\vert \psi _{L_{2}\mathbf{k}}\right\rangle \mapsto
\left\vert \psi _{L_{2}\mathfrak{M}\mathbf{k}}\right\rangle $ and $\left\vert \psi _{L_{1}\mathbf{k}}\right\rangle \leftrightarrow 
\left\vert \psi _{L_{3}\mathfrak{M}\mathbf{k}}\right\rangle $.

\subsection{Bare Hamiltonians} \label{H0_app}
\subsubsection{Hamiltonian at $\Gamma$}
We will write down the Hamiltonian at $\Gamma$ $\mathcal{H}_{\Gamma }$. Here it is easier to use a basis that makes the three-fold rotation
operator diagonal. The basis is unitarily transformed as $\tilde{\psi}%
_{\Gamma \mathbf{k}}^{\dagger }=\psi _{\Gamma \mathbf{k}}^{\dagger }\mathcal{%
U}$ and the symmetry operators transform, for instance, as $\mathcal{\tilde{C}}%
_{3,\Gamma }=\mathcal{U}^{-1}\mathcal{C}_{3,\Gamma }\mathcal{U}$, where the
unitary matrix 
\begin{equation}
\mathcal{U}=\left( 
\begin{array}{ccc}
\frac{1}{\sqrt{2}} & \frac{1}{\sqrt{2}} & 0 \\ 
\frac{i}{\sqrt{2}} & -\frac{i}{\sqrt{2}} & 0 \\ 
0 & 0 & 1%
\end{array}%
\right) .
\end{equation}%
In this ``tilde" basis, the symmetry operators are 
\begin{equation}
\begin{split}
\mathcal{\tilde{C}}_{3,\Gamma }=\left( 
\begin{array}{ccc}
e^{i\frac{2\pi }{3}} & 0 & 0 \\ 
0 & e^{-i\frac{2\pi }{3}} & 0 \\ 
0 & 0 & 1%
\end{array}%
\right) \text{, }\mathcal{\tilde{M}}_{\Gamma }=\left( 
\begin{array}{ccc}
0 & 1 & 0 \\ 
1 & 0 & 0 \\ 
0 & 0 & 1%
\end{array}%
\right) \text{, } 
\\
\mathcal{\tilde{I}}_{\Gamma }=\left( 
\begin{array}{ccc}
1 & 0 & 0 \\ 
0 & 1 & 0 \\ 
0 & 0 & -1%
\end{array}%
\right) \text{, }\mathcal{\tilde{T}}_{\Gamma }=\left( 
\begin{array}{ccc}
0 & 1 & 0 \\ 
1 & 0 & 0 \\ 
0 & 0 & 1%
\end{array}%
\right) K,
\end{split}
\end{equation}%
where the last operator is the time-reversal operator with complex conjugation $K$. This tilde Hamiltonian follows the symmetry constraints 
\begin{eqnarray}
\mathcal{\tilde{C}}_{3,\Gamma }\mathcal{\tilde{H}}_{\Gamma }(\mathbf{k})%
\mathcal{\tilde{C}}_{3,\Gamma }^{-1} &=&\mathcal{\tilde{H}}_{\Gamma }(%
\mathfrak{R}\mathbf{k}), \\
\mathcal{\tilde{I}}_{\Gamma }\mathcal{\tilde{H}}_{\Gamma }(\mathbf{k})%
\mathcal{\tilde{I}}_{\Gamma }^{-1} &=&\mathcal{\tilde{H}}_{\Gamma }(-\mathbf{%
k}), \\
\mathcal{\tilde{T}}_{\Gamma }\mathcal{\tilde{I}}_{\Gamma }\mathcal{\tilde{H}}%
_{\Gamma }(\mathbf{k})\left( \mathcal{\tilde{T}}_{\Gamma }\mathcal{\tilde{I}}%
_{\Gamma }\right) ^{-1} &=&\mathcal{\tilde{H}}_{\Gamma }(\mathbf{k}), \\
\mathcal{\tilde{M}}_{\Gamma }\mathcal{\tilde{H}}_{\Gamma }(\mathbf{k})%
\mathcal{\tilde{M}}_{\Gamma }^{-1} &=&\mathcal{\tilde{H}}_{\Gamma }(%
\mathfrak{M}\mathbf{k}).
\end{eqnarray}%
By taking $\mathcal{\tilde{H}}_{\Gamma }(\mathbf{k})=\left( 
\begin{array}{ccc}
\tilde{h}_{11} & \tilde{h}_{12} & \tilde{h}_{13} \\ 
\tilde{h}_{12}^{\ast } & \tilde{h}_{22} & \tilde{h}_{23} \\ 
\tilde{h}_{13}^{\ast } & \tilde{h}_{23}^{\ast } & \tilde{h}_{33}%
\end{array}%
\right) (\mathbf{k})$ and defining $k_{\pm }=k_{x}\pm ik_{y}$, symmetry
conditions require that for $\mathcal{\tilde{C}}_{3,\Gamma }$
\begin{widetext}
\begin{equation}
\left( 
\begin{array}{ccc}
\tilde{h}_{11} & \tilde{h}_{12} & \tilde{h}_{13} \\ 
& \tilde{h}_{22} & \tilde{h}_{23} \\ 
&  & \tilde{h}_{33}%
\end{array}%
\right) (e^{i\frac{2\pi }{3}}k_{+},e^{-i\frac{2\pi }{3}}k_{-},k_{z})=\left( 
\begin{array}{ccc}
\tilde{h}_{11} & e^{-i\frac{2\pi }{3}}\tilde{h}_{12} & e^{i\frac{2\pi }{3}}%
\tilde{h}_{13} \\ 
& \tilde{h}_{22} & e^{-i\frac{2\pi }{3}}\tilde{h}_{23} \\ 
&  & \tilde{h}_{33}%
\end{array}%
\right) (k_{+},k_{-},k_{z}),  \label{C3_H}
\end{equation}%
for $\mathcal{\tilde{I}}_{\Gamma }$%
\begin{equation}
\left( 
\begin{array}{ccc}
\tilde{h}_{11} & \tilde{h}_{12} & \tilde{h}_{13} \\ 
& \tilde{h}_{22} & \tilde{h}_{23} \\ 
&  & \tilde{h}_{33}%
\end{array}%
\right) (-k_{+},-k_{-},-k_{z})=\left( 
\begin{array}{ccc}
\tilde{h}_{11} & \tilde{h}_{12} & -\tilde{h}_{13} \\ 
& \tilde{h}_{22} & -\tilde{h}_{23} \\ 
&  & \tilde{h}_{33}%
\end{array}%
\right) (k_{+},k_{-},k_{z}),
\end{equation}%
for $\mathcal{\tilde{T}}_{\Gamma }\mathcal{\tilde{I}}_{\Gamma }$%
\begin{equation}
\left( 
\begin{array}{ccc}
\tilde{h}_{11} & \tilde{h}_{12} & \tilde{h}_{13} \\ 
& \tilde{h}_{22} & \tilde{h}_{23} \\ 
&  & \tilde{h}_{33}%
\end{array}%
\right) (k_{+},k_{-},k_{z})=\left( 
\begin{array}{ccc}
\tilde{h}_{22} & \tilde{h}_{12} & -\tilde{h}_{23}^{\ast } \\ 
& \tilde{h}_{11} & -\tilde{h}_{13}^{\ast } \\ 
&  & \tilde{h}_{33}%
\end{array}%
\right) (k_{+},k_{-},k_{z}),  \label{Mx_H}
\end{equation}%
and for $\mathcal{\tilde{M}}_{\Gamma }$ 
\begin{equation}
\left( 
\begin{array}{ccc}
\tilde{h}_{11} & \tilde{h}_{12} & \tilde{h}_{13} \\ 
& \tilde{h}_{22} & \tilde{h}_{23} \\ 
&  & \tilde{h}_{33}%
\end{array}%
\right) (-k_{-},-k_{+},k_{z})=\left( 
\begin{array}{ccc}
\tilde{h}_{22} & \tilde{h}_{12}^{\ast } & \tilde{h}_{23} \\ 
& \tilde{h}_{11} & \tilde{h}_{13} \\ 
&  & \tilde{h}_{33}%
\end{array}%
\right) (k_{+},k_{-},k_{z}),  \label{Mz_H}
\end{equation}%
respectively. The Hamiltonian up to second order is:
\begin{equation}
\mathcal{\tilde{H}}_{\Gamma }(\mathbf{k})=\left( 
\begin{array}{ccc}
ak_{||}^{2}+a^{\prime }k_{z}^{2}+\varepsilon _{1} & i ck_{-}k_{z}+c^{\prime
}k_{+}^{2} & dk_{+} \\ 
-ick_{+}k_{z}+c^{\prime }k_{-}^{2} & ak_{||}^{2}+a^{\prime
}k_{z}^{2}+\varepsilon _{1} & -dk_{-} \\ 
dk_{-} & -dk_{+} & bk_{||}^{2}+b^{\prime }k_{z}^{2}+\varepsilon _{2}%
\end{array}%
\right).
\end{equation}%
Here $k_{||}=k_{x}^{2}+k_{y}^{2}$ and all parameters $a$, $a^{\prime }$, $b$%
, $b^{\prime }$, $c$, $c^{\prime }$, $d$, $\varepsilon _{1}$, and $%
\varepsilon _{2}$ are real. Finally, we transform $\mathcal{\tilde{H}}%
_{\Gamma }$ back to $\mathcal{H}_{\Gamma }$ by%
\begin{eqnarray} \label{HGamma}
\mathcal{H}_{\Gamma }(\mathbf{k}) &=&\mathcal{U\tilde{H}}_{\Gamma }(\mathbf{k%
})\mathcal{U}^{-1} \\
&=&\left( 
\begin{array}{ccc}
ak_{||}^{2}+a^{\prime }k_{z}^{2}+\varepsilon _{1}+ck_{y}k_{z}+c^{\prime
}(k_{x}^{2}-k_{y}^{2}) & -ck_{x}k_{z}-2c^{\prime }k_{x}k_{y} & i\sqrt{2}dk_{y}
\\ 
-ck_{x}k_{z}-2c^{\prime }k_{x}k_{y} & ak_{||}^{2}+a^{\prime
}k_{z}^{2}+\varepsilon _{1}-ck_{y}k_{z}-c^{\prime }(k_{x}^{2}-k_{y}^{2}) & i%
\sqrt{2}dk_{x} \\ 
-i\sqrt{2}dk_{y} & -i\sqrt{2}dk_{x} & bk_{||}^{2}+b^{\prime
}k_{z}^{2}+\varepsilon _{2}%
\end{array}%
\right) ,  \notag  
\end{eqnarray}
\end{widetext}

\subsubsection{Hamiltonians at $L$ and $M$}
The Hamiltonian at $L$'s $\mathcal{H}_{L}$ is identical to $\mathcal{H}_{M}$, and can be written as
\begin{equation}
\mathcal{H}_{L}(\mathbf{k})=\left( 
\begin{array}{ccc}
\xi _{1} & 0 & 0 \\ 
0 & \xi _{2} & 0 \\ 
0 & 0 & \xi _{3}%
\end{array}%
\right) (\mathbf{k}).
\end{equation}%
Here we assume that within the low-energy region, the three bands are independent. The symmetry operators are then given by
\begin{equation}
\begin{split}
\mathcal{C}_{3,L}=\left( 
\begin{array}{ccc}
0 & 0 & 1 \\ 
1 & 0 & 0 \\ 
0 & 1 & 0%
\end{array}%
\right) \text{, }\mathcal{M}_{L}=\left( 
\begin{array}{ccc}
0 & 0 & 1 \\ 
0 & 1 & 0 \\ 
1 & 0 & 0%
\end{array}%
\right) \text{, }
\\
\mathcal{I}_{L}=\left( 
\begin{array}{ccc}
1 & 0 & 0 \\ 
0 & 1 & 0 \\ 
0 & 0 & 1%
\end{array}%
\right) \text{, }\mathcal{T}_{L}=\left( 
\begin{array}{ccc}
1 & 0 & 0 \\ 
0 & 1 & 0 \\ 
0 & 0 & 1%
\end{array}%
\right) K.
\end{split}
\end{equation}%
Similar as before, we conclude that 
\begin{eqnarray}
\xi _{2}(\mathbf{k}) &=&A k_{x}^{2}+B k_{y}^{2}+C k_{z}^{2}+D
k_{y}k_{z}+\varepsilon _{3}, \\
\xi _{1}(\mathbf{k}) &=&\xi _{2}(\mathfrak{R}\mathbf{k}),~\xi _{3}(\mathbf{k})=\xi _{2}(\mathfrak{R}^{-1}\mathbf{k}).
\end{eqnarray}

\begin{table}[tbp]
\caption{Character table for the point group $D_{3d}$. Only $E_{g}$ and $%
A_{2u}$ IRs for the three valence bands are shown.}
\label{table:D3d}
\begin{center}
\setlength{\tabcolsep}{1.2em} 
\begin{tabular}{c|cccccc}
\hline
$D_{3d}$ & $E$ & $2C_{3}$ & $3C_{2}$ & $i$ & $2S_{6}$ & $\sigma _{h}$ \\ 
\hline
$E_{g}$ & $2$ & $-1$ & $0$ & $2$ & $-1$ & $0$ \\ 
$A_{2u}$ & $1$ & $1$ & $-1$ & $-1$ & $-1$ & $1$ \\ \hline
\end{tabular}%
\end{center}
\end{table}

\subsubsection{Hamiltonians at $A$}
Although the conduction band ($A_g$) extends monotonically along $L$-$M$, in some calculations the band structure shows a band inversion along $\Gamma$-$A$, where the $E_g$ valence bands switch with the $E_u$ conduction bands. We consider this case in which the valence band becomes IR $E_u$ at $A$.

The $E_u$ bands are two-fold degenerate at $A$, where they are described by the same Hamiltonian (with different parameters) as the upper $2 \times 2$ block of $\mathcal{H}_{\Gamma}$ in Eq. (\ref{HGamma}). In addition, we show the coupling Hamiltonian between the $E_g$ and the $E_u$ bands in $\mathcal{V}$. In the basis where the symmetry operators for the $E_u$ bands take the form
\begin{equation}
\mathcal{I}'_{\Gamma } =\left( 
\begin{array}{cc}
-1 & 0 \\ 
0 & -1
\end{array}%
\right),~ 
\mathcal{M}'_{\Gamma } =\left( 
\begin{array}{cc}
1 & 0 \\ 
0 & -1
\end{array}%
\right) ,~
\mathcal{C}'_{3,\Gamma } = \left( 
\begin{array}{cc}
-\frac{1}{2} & \frac{\sqrt{3}}{2} \\ 
-\frac{\sqrt{3}}{2} & -\frac{1}{2} \\ 
\end{array}%
\right) ,
\end{equation}
the $k \cdot p$ model will read
\begin{equation}
\mathcal{H}_{\Gamma A} =  \left( 
\begin{array}{cc}
t_1 k_z & i t_2 k_x k_y k_z  \\ 
-i t_2 k_x k_y k_z  & t_1 k_z  \\ 
\end{array}%
\right)
\end{equation}
with real parameters $t_1$ and $t_2$. The coupling Hamiltonian indicates anticrossing among the $E_g$ and $E_u$ bands.

\subsection{CDW gap functions} \label{CDWgap_app}

The gap functions in Eqs. (\ref{symmetry1}-\ref{symmetry3}) follow 
\begin{eqnarray}
\mathcal{C}_{3,\Gamma }\mathcal{D}(\mathbf{k})\mathcal{C}_{3,L}^{-1} &=&%
\mathcal{D}(\mathfrak{R}\mathbf{k}),  \label{gap_s1} \\
\mathcal{I}_{\Gamma }\mathcal{D}(\mathbf{k})\mathcal{I}_{L}^{-1} &=&\eta _{I}%
\mathcal{D}(\mathfrak{-}\mathbf{k}),  \label{gap_s2} \\
\mathcal{M}_{\Gamma }\mathcal{D}(\mathbf{k})\mathcal{M}_{L}^{-1} &=&\eta _{M}%
\mathcal{D}(\mathfrak{M}\mathbf{k}),  \label{gap_s3}
\end{eqnarray}%
where $\eta _{C_{2}},~\eta _{M}$ are eigenvalues of the corresponding
operations for the OP. The allowed values for $\eta _{I}$ and $\eta _{M}$
are $\pm 1$. The four IRs $L_{1}^{+},~L_{2}^{+},~L_{1}^{-},~L_{2}^{-}$ have symmetries with $(\eta _{I},\eta _{M})=(1,1),~(1,-1),~(-1,-1),~(-1,1)$,
respectively. To obtain the gap functions, we have, through Eqs. (\ref{gap_s2}) and (\ref{gap_s3}),
\begin{widetext}
\begin{equation}
\left( 
\begin{array}{ccc}
\Delta _{11} & \Delta _{12} & \Delta _{13} \\ 
\Delta _{21} & \Delta _{22} & \Delta _{23} \\ 
\Delta _{31} & \Delta _{32} & \Delta _{33}%
\end{array}%
\right) (\mathbf{k}) 
 =\eta _{I}\left( 
\begin{array}{ccc}
\Delta _{11} & \Delta _{12} & \Delta _{13} \\ 
\Delta _{21} & \Delta _{22} & \Delta _{23} \\ 
-\Delta _{31} & -\Delta _{32} & -\Delta _{33}%
\end{array}%
\right) (-\mathbf{k}) 
=\eta _{M}\left( 
\begin{array}{ccc}
\Delta _{13} & \Delta _{12} & \Delta _{11} \\ 
-\Delta _{23} & -\Delta _{22} & -\Delta _{21} \\ 
\Delta _{33} & \Delta _{32} & \Delta _{31}%
\end{array}%
\right) (\mathfrak{M}\mathbf{k}).
\end{equation}%

In particular, the gap functions associated with $L_{2}$, which is invariant under both inversion and the mirror symmetry, are: 
\begin{align}
\begin{split}
\Delta _{12}(\mathfrak{-}\mathbf{k}) &=\eta _{I}\Delta _{12}(\mathbf{k}%
),~\Delta _{22}(\mathfrak{-}\mathbf{k})=\eta _{I}\Delta _{22}(\mathbf{k}%
),~
\\
\Delta _{32}(\mathfrak{-}\mathbf{k}) &=-\eta _{I}\Delta _{32}(\mathbf{k}),
\end{split}
\\
\begin{split}
\Delta _{12}(\mathfrak{M}\mathbf{k}) &=\eta _{M}\Delta _{12}(\mathbf{k}%
),~\Delta _{22}(\mathfrak{M}\mathbf{k})=-\eta _{M}\Delta _{22}(\mathbf{k}%
),~
\\
\Delta _{32}(\mathfrak{M}\mathbf{k})&=\eta _{M}\Delta _{32}(\mathbf{k}).
\end{split}
\end{align} 
So the gap functions for the four CDW states to lowest order in $k$ are: 
\begin{align}
L_{1}^{+} &:~\Delta _{12}(\mathbf{k}) \sim \mathrm{const.},~\Delta _{22}(%
\mathbf{k})\sim \lbrace k_{x}k_{y},k_{x} k_{z} \rbrace,~
\Delta _{32}(\mathbf{k}) \sim
\lbrace k_{y},k_{z} \rbrace ,~ 
\\
L_{2}^{+} &:~\Delta _{12}(\mathbf{k}) \sim \lbrace k_{x}k_{y},k_{x}k_{z} \rbrace ,~\Delta _{22}(%
\mathbf{k})\sim \mathrm{const.},~
\Delta _{32}(\mathbf{k}) \sim k_{x},
\\
L_{1}^{-}& :~\Delta _{12}(\mathbf{k})\sim k_{x},~\Delta _{22}(\mathbf{k}%
)\sim \lbrace k_{y},k_{z} \rbrace ,~\Delta _{32}(\mathbf{k})\sim \lbrace k_{x}k_{y}, k_{x}k_{z} \rbrace ,~ \\
L_{2}^{-}& :~\Delta _{12}(\mathbf{k})\sim \lbrace k_{y},k_{z} \rbrace ~\Delta _{22}(\mathbf{%
k})\sim k_{x},~\Delta _{32}(\mathbf{k})\sim \mathrm{const.}
\end{align}
Two functions in the brackets are assumed to combine linearly.
The remaining gap functions can be obtained via the three-fold rotations as follows.
From Eq. (\ref{gap_s1}),

\begin{equation}
\left( 
\begin{array}{ccc}
\Delta _{11} & \Delta _{12} & \Delta _{13} \\ 
\Delta _{21} & \Delta _{22} & \Delta _{23} \\ 
\Delta _{31} & \Delta _{32} & \Delta _{33}%
\end{array}%
\right) (\mathfrak{R}\mathbf{k})=\left( 
\begin{array}{ccc}
-\frac{1}{2}\Delta _{13}+\frac{\sqrt{3}}{2}\Delta _{23} & -\frac{1}{2}\Delta
_{11}+\frac{\sqrt{3}}{2}\Delta _{21} & -\frac{1}{2}\Delta _{12}+\frac{\sqrt{3%
}}{2}\Delta _{22} \\ 
-\frac{1}{2}\Delta _{23}-\frac{\sqrt{3}}{2}\Delta _{13} & -\frac{1}{2}\Delta
_{21}-\frac{\sqrt{3}}{2}\Delta _{11} & -\frac{1}{2}\Delta _{22}-\frac{\sqrt{3%
}}{2}\Delta _{12} \\ 
\Delta _{33} & \Delta _{31} & \Delta _{32}%
\end{array}%
\right) (\mathbf{k}).
\end{equation}%

Therefore, 
\begin{equation}
\begin{split}
\left( 
\begin{array}{c}
\Delta _{11} \\ 
\Delta _{21}%
\end{array}%
\right) (\mathbf{k})& =\left( 
\begin{array}{cc}
-\frac{1}{2} & -\frac{\sqrt{3}}{2} \\ 
\frac{\sqrt{3}}{2} & -\frac{1}{2}%
\end{array}%
\right) \left( 
\begin{array}{c}
\Delta _{12} \\ 
\Delta _{22}%
\end{array}%
\right) (\mathfrak{R}\mathbf{k}), \\
\left( 
\begin{array}{c}
\Delta _{13} \\ 
\Delta _{23}%
\end{array}%
\right) (\mathbf{k})& =\left( 
\begin{array}{cc}
-\frac{1}{2} & \frac{\sqrt{3}}{2} \\ 
-\frac{\sqrt{3}}{2} & -\frac{1}{2}%
\end{array}%
\right) \left( 
\begin{array}{c}
\Delta _{12} \\ 
\Delta _{22}%
\end{array}%
\right) (\mathfrak{R}^{-1}\mathbf{k}), \\
\Delta _{31}(\mathbf{k})& =\Delta _{32}(\mathfrak{R}\mathbf{k}),~\Delta
_{33}(\mathbf{k})=\Delta _{32}(\mathfrak{R}^{-1}\mathbf{k}).
\end{split}
\label{Rd}
\end{equation}
\end{widetext}

For the CDW gap functions among the $E_u$ valence bands at $A$ and the conduction bands at $M$'s, an analysis along the preceding lines yields: 
\begin{align}
L_{1}^{+}& :~\Delta _{42}(\mathbf{k})\sim \lbrace k_{y}, k_{z} \rbrace,~\Delta _{52}(%
\mathbf{k})\sim k_x, ~ \\
L_{2}^{+}& :~\Delta _{42}(\mathbf{k})\sim k_{x} ,~\Delta _{52}(%
\mathbf{k})\sim ,  \lbrace k_{y}, k_{z} \rbrace \\
L_{1}^{-}& :~\Delta _{42}(\mathbf{k})\sim \lbrace k_{x} k_{y},k_{x} k_{z} \rbrace,~\Delta _{52}(\mathbf{k}%
)\sim \mathrm{const.} ,~\\
L_{2}^{-}& :~\Delta _{42}(\mathbf{k})\sim \mathrm{const.}, ~\Delta _{52}(\mathbf{%
k})\sim \lbrace k_{x} k_{y},k_{x} k_{z} \rbrace.
\end{align}
where the momentum $\mathbf{k}$ is relative to the $A$ point and the subscript $4,~5$ are used to denote the two $E_u$ bands.

\section{Ginzburg-Landau theory for the two order parameters} \label{GL_app}
We examine the Ginzburg-Landau theory for primary and secondary order parameters (OPs). The primary OP $\vec{\phi} =  \left(\phi_1,\phi_2,\phi_3\right)$ is taken to belong to the $L_{1}^{-}$ irreducible representation (IR), which is delineated by 
\begin{equation}
L_{1}^{-} =A_{u}\otimes \left( e^{-i\mathbf{Q}_{1}\cdot \mathbf{T}},e^{-i \mathbf{Q}_{2}\cdot \mathbf{T}},e^{-i\mathbf{Q}_{3}\cdot \mathbf{T}}\right),
\end{equation}
where $A_{u}$ is an IR in the point group $C_{2h}$ (little co-group at $L$) and $e^{-i\mathbf{Q}_{i}\cdot \mathbf{T}}$ stand for translational representations. The 3D ordering vectors are defined by $\mathbf{Q}_{1}=\frac{1}{2} (\mathbf{a}_1^*+ \mathbf{a}_3^*)$, $\mathbf{Q}_{2}=\frac{1}{2} (\mathbf{a}_2^*+ \mathbf{a}_3^*)$, and $\mathbf{Q}_{3}=\frac{1}{2} (\mathbf{a}_1^*+\mathbf{a}_2^*+ \mathbf{a}_3^*)$.
The images of $\vec{\phi}$ under symmetry operations in the space group are
\begin{equation}
\begin{split} 
    3_{[001]} : {}& ~ \left(\phi_1,\phi_2,\phi_3\right) \rightarrow \left(\phi_2,\phi_3,\phi_1\right), \\
    m_{[100]}:{}& ~ \left(\phi_1,\phi_2,\phi_3\right) \rightarrow \left(-\phi_3,-\phi_2,-\phi_1\right), \\
    \bar{I} :{}& ~  \left(\phi_1,\phi_2,\phi_3\right) \rightarrow \left(-\phi_1,-\phi_2,-\phi_3\right), 
\\
    T_{\mathbf{a}_1} :{}& ~ \left(\phi_1,\phi_2,\phi_3\right) \rightarrow \left(-\phi_1,\phi_2,-\phi_3\right),  \\
T_{\mathbf{a}_2} :{}& ~ \left(\phi_1,\phi_2,\phi_3\right) \rightarrow \left(\phi_1,-\phi_2,-\phi_3\right),  \\
T_{\mathbf{a}_3} :{}& ~ \left(\phi_1,\phi_2,\phi_3\right) \rightarrow \left(-\phi_1,-\phi_2,-\phi_3\right).  
\end{split} \label{sym_primaryOP}
\end{equation}
Translations by $\mathbf{a}_{1,2,3}$ will produce minus signs due to the oscillating nature of the CDW. From the collection of these images, one realizes that by treating $\vec{\phi}$ as an ordinary coordinate vector $\vec{x}=\left( x,y,z \right)$, the transformations correspond to those in the point group $T_{h}$. In other words, this is a homomorphism: the space group generates in the vector space of $\vec{\phi}$ the point group $T_{h}$. For other three IRs ($L_{1}^{+}, L_{2}^{+}, L_{2}^{-}$), it is straightforward to obtain the associated images, which are similar to those of $L_{1}^{-}$ with modifications at $m_{[100]}$ and $\bar{I}$. Their image groups are also $T_{h}$.

The secondary OP $\vec{\zeta}$ also transforms as the $M_{1}^{+}$ IR: 
\begin{equation}
M_{1}^{+} =A_{g}\otimes \left( e^{-i\mathbf{Q}'_{1}\cdot \mathbf{T}},e^{-i \mathbf{Q}'_{2}\cdot \mathbf{T}},e^{-i\mathbf{Q}'_{3}\cdot \mathbf{T}}\right).
\end{equation}
The secondary OP has the $A_{g}$ symmetry and importantly it describes a 2D CDW with ordering vectors $\mathbf{Q}'_{1}=\frac{1}{2} \mathbf{a}_1^*$, $\mathbf{Q}'_{2}=\frac{1}{2} \mathbf{a}_2^*$, and $\mathbf{Q}'_{3}=\frac{1}{2}(\mathbf{a}_1^*+\mathbf{a}_2^*)$. The images of $\vec{\zeta}$ under the symmetry operations  are
\begin{equation}
\begin{split}
3_{[001]} : {}& ~ \left(\zeta_1,\zeta_2,\zeta_3\right) \rightarrow \left(\zeta_2,\zeta_3,\zeta_1\right), \\
m_{[100]} :{}&  ~ \left(\zeta_1,\zeta_2,\zeta_3\right) \rightarrow \left(\zeta_3,\zeta_2,\zeta_1\right),  \\
\bar{I} :{}&  ~ \left(\zeta_1,\zeta_2,\zeta_3\right) \rightarrow \left(\zeta_1,\zeta_2,\zeta_3\right),  \\
T_{\mathbf{a}_1} :{}& ~ \left(\zeta_1,\zeta_2,\zeta_3\right) \rightarrow \left(-\zeta_1,\zeta_2,-\zeta_3\right),  \\
T_{\mathbf{a}_2} :{}& ~ \left(\zeta_1,\zeta_2,\zeta_3\right) \rightarrow \left(\zeta_1,-\zeta_2,-\zeta_3\right),  \\
T_{\mathbf{a}_3} :{}&  ~ \left(\zeta_1,\zeta_2,\zeta_3\right) \rightarrow \left(\zeta_1,\zeta_2,\zeta_3\right).
\end{split}
\end{equation}
Note that a translation in $\mathbf{a}_3$ has no effect due to the 2D character. Now the space group generates the point group $T$ for $\vec{\zeta}$. Incidentally, the image group for $M_{1}^{+}$ IR is group $T$ and is $T_h$ for both $M_{1}^{-}$ and $M_{2}^{-}$ IRs.

The Ginzburg-Landau free energy for the primary OP $\vec{\phi}$ is invariant for the group $T_{h}$ and it can be written as:
\begin{equation}
F_{\mathrm{primary}} =\frac{1}{2}\alpha \vec{\phi}^2 + \frac{1}{4}\beta_{1} \vec{\phi}^4 + \frac{1}{4}\beta_{2} \left( \phi_1^4 + \phi_2^4 + \phi_3^4 \right) + \Delta F_{\mathrm{primary}},  \label{Fprimary}
\end{equation} 
where $\vec{\phi}^2=\phi_1^2+\phi_2^2 + \phi_3^2$ is defined. It is known that to have a conclusive phase diagram, sixth-degree terms, 
\begin{equation}
\begin{split}
\Delta F_{\mathrm{primary}} &=- \frac{1}{6}\gamma_1 \vec{\phi}^6 + \frac{1}{2} \gamma_2 \vec{\phi}^2 \left( \phi_1^4 + \phi_2^4 + \phi_3^4 \right) \\
&+ \frac{1}{2} \gamma_3 \phi_1^2 \phi_2^2 \phi_3^2 + \frac{1}{2} \gamma_4 \left( \phi_1^2 - \phi_2^2 \right) \left( \phi_1^2 - \phi_3^2 \right) \left( \phi_3^2 - \phi_1^2 \right),  \label{Fprimary2}
\end{split}
\end{equation} 
are required \cite{Toledano,Gufan1973}. When $T<T_c$ and $\alpha<0$, a CDW state for nonzero $\vec{\phi}$ becomes the ground state. Moreover, only the so-called first type of CDW phases are possible~\cite{Toledano,Gufan1973}: either of $\vec{\phi}_{\mathrm{I}}\propto\left(1,1,1 \right)$, $\vec{\phi}_{\mathrm{II}}\propto\left(1,1,0 \right)$, or $\vec{\phi}_{\mathrm{III}} \propto\left(1,0,0 \right)$. The free energies for these three phases are given by
\begin{equation}
F_{\mathrm{primary},M} = -\frac{1}{48 D_{M}} \left(s_{M} - t_{M} \right)^{2} \left( s_{M} +2 t_{M} \right),
\end{equation}
where $M=\mathrm{I},~\mathrm{II},~\mathrm{III}$, and
\begin{align}
D_{\mathrm{I}} = \gamma_{3} + \gamma_2 + \frac{1}{9}\gamma_3,~
t_{\mathrm{I}} = \beta_{1} + \frac{1}{3} \beta_{2},~ s_{\mathrm{I}} = \sqrt{ t_{\mathrm{I}}^2 -4\alpha D_{\mathrm{I}}}, \\
D_{\mathrm{II}} = \gamma_{1} + 3 \gamma_2 ,~
t_{\mathrm{II}} = \beta_{1} + \beta_{2},~ s_{\mathrm{II}} = \sqrt{ t_{\mathrm{II}}^2 -4\alpha D_{\mathrm{II}}}, \\
D_{\mathrm{III}} = \gamma_{1} + \frac{3}{2} \gamma_2 ,~
t_{\mathrm{III}} = \beta_{1} + \frac{1}{2} \beta_{2},~ s_{\mathrm{III}} = \sqrt{ t_{\mathrm{III}}^2 -4\alpha D_{\mathrm{III}}}. \\
\end{align}
An analysis of the first-order transitions between these phases is outside of scope of this paper. We will assume that the $M=\mathrm{I}$ phase is the favorable one for given parameters.

The free energy for the secondary OP is 
\begin{equation}
F_{\mathrm{secondary}} =\frac{1}{2} \alpha' \vec{\zeta}^2 +\frac{1}{4} \beta'_{1} \vec{\zeta}^4 + \frac{1}{4} \beta'_{2} \left( \zeta_1^4 + \zeta_2^4 + \zeta_3^4 \right) + \gamma_1' \zeta_1 \zeta_2 \zeta_3. 
\end{equation}
The expansion of the free energy to the fourth degree is sufficient for our discussion. Because the vector space of $\vec{\zeta}$ has group $T$ symmetry, a cubic term is observed, and the lowest-degree coupling form reads as
\begin{equation}
F_{\mathrm{coupling}} = \frac{1}{3} \lambda \left( \phi_1 \phi_2 \zeta_3 + \phi_2 \phi_3 \zeta_1 + \phi_3 \phi_1 \zeta_2\right),
\end{equation}
in which the primary OP is quadratic and the secondary OP is linear. 

The coupling with linear $\zeta$ indicates that the secondary OP will coincide with the primary OP. To see this, we reduce the problem by taking $\phi=\phi_1=\phi_2=\phi_3$ and $\zeta= \zeta_1= \zeta_2= \zeta_3$, omit the sixth degree terms and obtain the total free energy as
\begin{equation}
F =\frac{1}{2} \alpha \phi^2 + \frac{1}{4} \beta \phi^4 +\frac{1}{2} \alpha' \zeta^2 + \frac{1}{3} \gamma' \zeta^3 + \frac{1}{4} \beta' \zeta^4 + \lambda \zeta \phi^2. 
\end{equation}
The optimal values of $\phi$ and $\zeta$ ($\bar{\phi}$ and $\bar{\zeta}$) are obtained from $\frac{\partial F}{\partial \phi} \vert _{\bar{\phi},\bar{\zeta}}= \frac{\partial F}{\partial \zeta}\vert _{\bar{\phi},\bar{\zeta}} =0$ that give
\begin{align}
\bar{\phi}^2 = \frac{1}{\beta} \left( -\alpha - 2\lambda \bar{\zeta} \right), \label{phi_op} \\
\beta' \bar{\zeta}^3 + \gamma' \bar{\zeta}^2 + \alpha' \bar{\zeta} + \lambda \bar{\phi}^2 =0. \label{zeta_op}
\end{align}
Based on the fact that the secondary OP is induced by the primary OP, we demand that $\bar{\zeta}=0$ when $\bar{\phi} =0$, suggesting that $ \alpha' >0$ as well as $\gamma'^{2}-4\alpha' \beta'<0$. 

For nonzero $\bar{\phi} $, it happens when $\alpha + 2\lambda \bar{\zeta}<0$, which indicates that $T_{c}$ is modified by the nonzero of the secondary OP. 
Replacing $\bar{\phi}$ in Eq. (\ref{zeta_op}) by Eq. (\ref{phi_op}) yields the cubic equation,
\begin{equation}
\beta' \bar{\zeta}^3 +\gamma' \bar{\zeta}^2 + \left( \alpha' -  \frac{2 \lambda^2}{\beta} \right) \bar{\zeta} - \frac{\lambda \alpha}{\beta} =0.
\end{equation}
If the constant term $\lambda \alpha /\beta$ is finite, the cubic equation always has a real root, explaining the coincidence of the primary and secondary OPs. Close to $T_c$ when OPs are small, $\bar{\zeta} \propto \bar{\phi}^2 $ and thus $\bar{\psi} \propto \bar{\phi}^3 $. 

\section{Computational Details } 

The first-principles calculations were performed with the projector augmented wave method within the DFT~\cite{dft} framework as implemented in the VASP code~\cite{paw,vasp}. The GGA with the Perdew-Burke-Ernzerhof parameterization and modified Becke-Johnson (mBJ) meta-GGA were used to include the exchange-correlation effects~\cite{gga,mBJ}. We used a plane wave energy cut-off of 380 eV and a $\Gamma$-centered $ 12 \times12\times 8$ k-mesh to sample the bulk Brillouin zone of 1$T$-TiSe$_2$. To access the nodal lines in the full Brillouin zone, we constructed a tight-binding model Hamiltonian by projecting first-principles results onto Wannier orbitals using the VASP2WANNIER90 interface~\cite{vasp,wannier90}. The Ti {\it d} and Se {\it p} states were included in the construction of Wannier functions.

\section{Band structure of the L1$^-$ CDW state and nodal lines } \label{bands_kz}

\begin{figure*}
\centering
\includegraphics[width=0.9\textwidth]{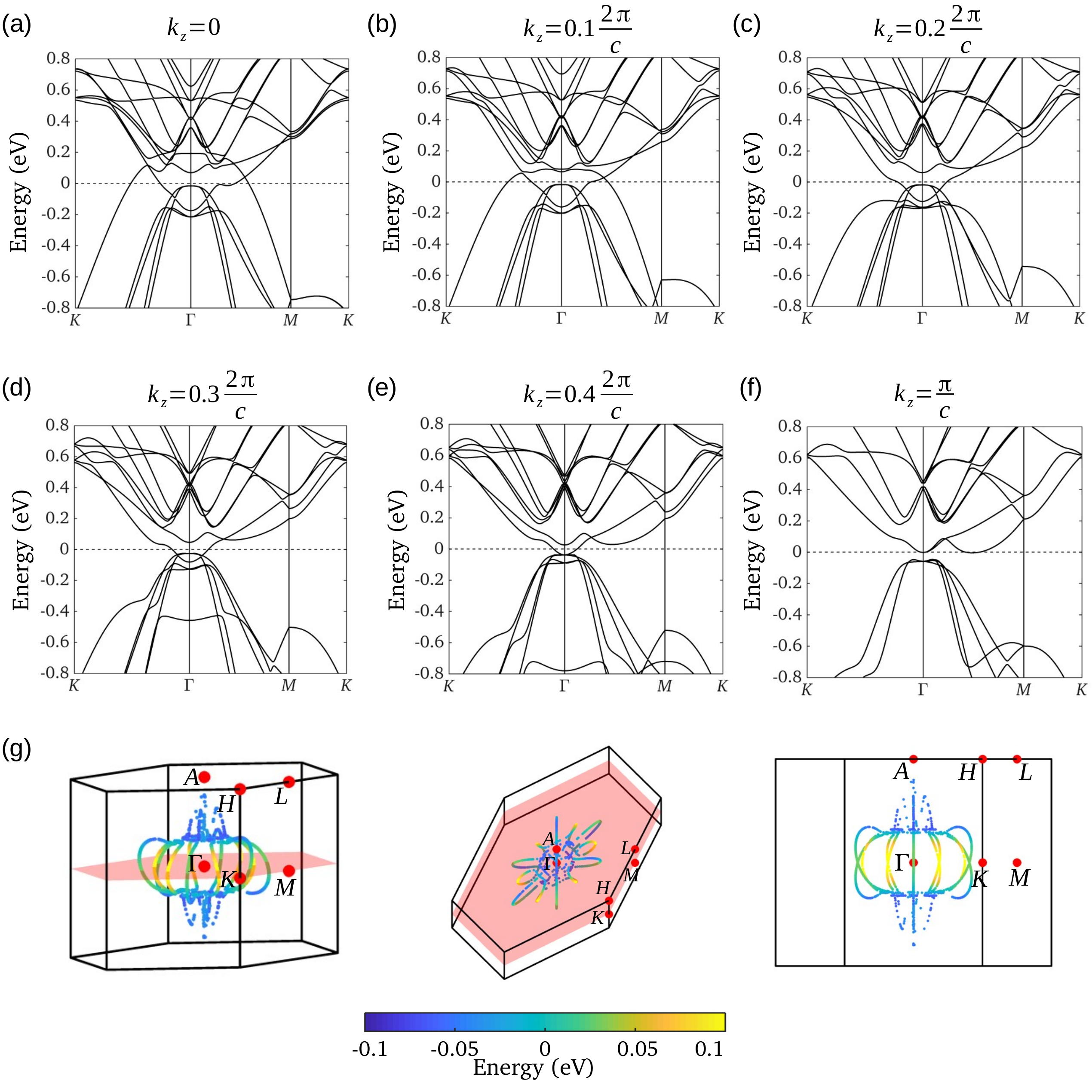}
\caption{Bulk structure of the $L_{1}^{-}$ CDW state with PLD at the selected $k_z$ planes. (a) $k_z=0$, (b) $k_z=0.1\frac{2\pi}{c}$, (c) $k_z=0.2\frac{2\pi}{c}$, (d) $k_z=0.3\frac{2\pi}{c}$, (e) $k_z=0.4\frac{2\pi}{c}$, and (f) $k_z=\frac{\pi}{c}$. The bulk nodal band crossings are absent at the $k_z=\frac{\pi}{c}$ plane. (g) Calculated nodal-lines crossings in the BZ. Light red plane marks the $k_z=0$ plane. The color scale highlights the nodal band crossing energy relative to the Fermi energy.}
\label{fig:nodal}s
\end{figure*}

We present the {\it ab-initio} band structure of the CDW state on various $k_z $ planes in Figs.~\ref{fig:nodal}(a)-(f). The band structure in Fig. \ref{fig:nodal}(a) shows the states at the $k_z = 0$ plane, which is the same band structure that is presented in the main text. The bands are seen to cross at discrete points along the $\Gamma-K$ and $\Gamma-M$ directions at an energy $\sim +0.1$ eV above the Fermi level. With an increase in the $k_z$ values, the band crossings start lowering their energies and finally merge into the $\Gamma$-point on the $k_z \sim 0.4\frac{2\pi}{c}$ plane at an energy $\sim -0.05$ eV, see Fig. \ref{fig:nodal}(e). On further increasing the $k_z$ value to $\frac{\pi}{c}$, the bands become gapped with a clear band gap in the band structure. A full BZ exploration of the band structure shows that the nodal points trace out the continuous lines which pass through the $k_z=0$ plane as shown in Fig. \ref{fig:nodal}(g). We find a total of 12 nodal lines which cross along the $\Gamma-A$ direction in the BZ. These results are in accord with the theoretical analysis presented above and in the main text. 

\section{Band structures under pressure}  \label{bands_pressure}

\begin{figure}
\centering
\includegraphics[width=0.5\textwidth]{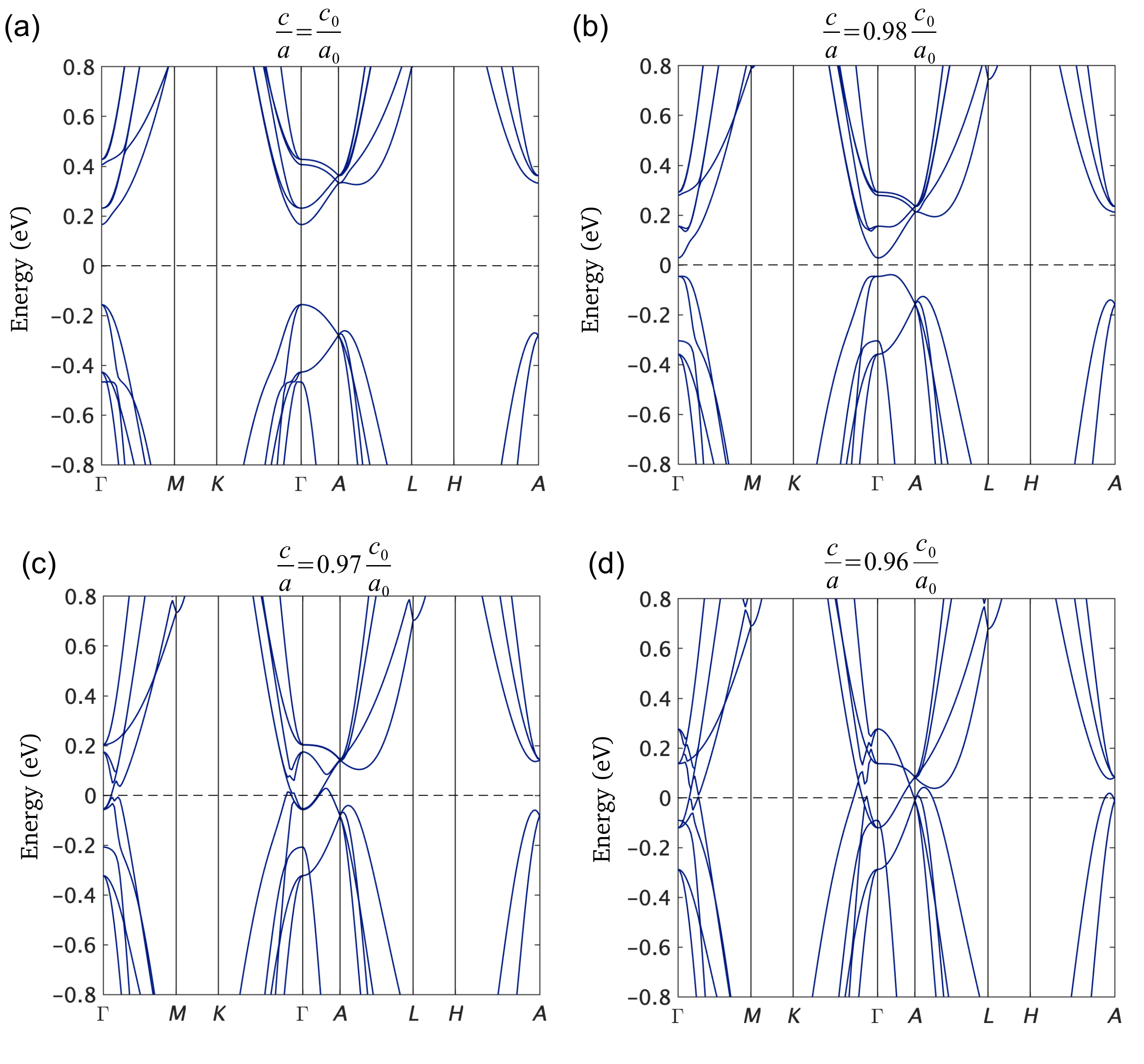}
\caption{The bulk band structure of a $2\times 2 \times 2$ superlattice of 1\textit{T}-TiSe$_2$ under hydrostatic pressure. (a) $c/a=c_0/a_0$, (b) $c/a=0.98 c_0/a_0$, (c) $c/a=0.97 c_0/a_0$, and (d) $c/a=0.96 c_0/a_0$. A CDW insulator to a normal metal transition is seen to take place with pressure. }
\label{fig:bandsmBJ}
\end{figure}

Figure~\ref{fig:bandsmBJ} shows the band structure of 1$T$-TiSe$_2$ $2\times 2 \times 2$ superlattice obtained with mBJ meta-GGA for various hydrostatic pressures. The band structure for $c/a=c_0/a_0$ [$c$ and $a$ ($c_0$ and $a_0$) are the lattice constants of pressurized (pristine) $2\times 2 \times 2$ 1\textit{T}-TiSe$_2$] is insulating with a well-defined gap between the valence and conduction bands. Due to this bandgap, the $A_{2u}$ band at $\Gamma$ does not show a band inversion. As we decrease $c/a$ to apply a hydrostatic pressure, the bandgap decreases and finally vanishes for $c/a=0.97 c_0/a_0$, realizing a metallic state. There is a band inversion between valence and conduction bands, realizing multiple nodal crossings for $c/a=0.96 c_0/a_0$. With further lowering of the lattice parameters, the overlap between the valence and conduction bands increases without showing any band hybridizations. This implies that the system is metallic without any CDW order at high pressures. 

\bibliographystyle{apsrev4-1} 
\bibliography{TiSe2_ref}

\end{document}